\documentclass[a4paper,10pt,twoside]{cpc-hepnp}

\usepackage{multicol}
\usepackage{graphicx}
\usepackage{booktabs}
\usepackage{amssymb,bm,mathrsfs,bbm,amscd}
\usepackage[tbtags]{amsmath}
\usepackage{lastpage}
\usepackage{color}

\begin{document}

\fancyhead[c]{\small Physics of Dark Universe~~~Vol. xx, No. x (2019) xxxxxx}
\fancyfoot[C]{\small 1-\thepage}

\footnotetext[0]{Submitted Oct. 2019}

\title{Buchdahl model in $f(R, T)$ gravity: A comparative study  with standard Einstein's gravity}

\author{$^1${S. K. Maurya}\email{sunil@unizwa.edu.om}, $^2${Ayan Banerjee}\email{ayan7575@yahoo.co.in}, $^3${Francisco Tello-Ortiz}\email{francisco.tello@ua.cl}
}
\maketitle

\address{$^1$ Department of Mathematical \& Physical Sciences, College of Arts \& Science, University of Nizwa, Nizwa, Sultanate of Oman\\
$^2$Astrophysics and Cosmology Research Unit, University of KwaZulu Natal, Private Bag X54001, Durban 4000,
South Africa\\
$^3$Departamento de F\'isica,
Facultad de ciencias b\'asicas, Universidad de Antofagasta, Casilla 170, Antofagasta, Chile}

\begin{abstract}
This paper is devoted in the study of the hydrostatic equilibrium of stellar structure in the framework of modified  $f(R, T)$  gravity theory that allows the non-conservation of energy-momentum, with possible implications for several cosmological and astrophysical issues such as the late-time cosmic acceleration of the universe without appealing to exotic matter fields. For this purpose, we consider the gravitational Lagrangian by taking an arbitrary function of the Ricci scalar and the trace of the stress-energy tensor. We obtain a generic form for the gravitational field equations and derive the field equation for $f(R,T)$  = $R+2 \chi T$. Here we propose a particular metric potential \textit{Buchdahl ansatz} [Phys. Rev. D 116, 1027 (1959)] in principle, of explaining almost all the known analytic solutions to the spherically symmetric, static Einstein equations with a perfect fluid source. For the choice of $f(T)= 2\chi T$ one may observe that the pressure and energy density profiles are markedly different. Important cases, which have been analyzed in detail, are all possible Buchdahl solutions for spherical equilibrium configuration in $f(R, T)$ gravity and compare them with standard gravity theory. We find that Buchdahl's solution in Einstein gravity and $f(R,T)$ gravity behaves in a similar manner but in some situations Einstein gravity displays more pleasing behavior than its $f(R,T)$ counterpart.  
\end{abstract}



\footnotetext[0]{\hspace*{-3mm}\raisebox{0.3ex}{$\scriptstyle\copyright$}2019
Particle Astrophysics and Cosmology, Elsevier Publication.}%

\begin{multicols}{2}

\maketitle

\section{Introduction}\label{sec1}
Over the past decade, the late-time acceleration of the universe has led to new perspectives and scenarios in the field of modern cosmology and physics as a whole. Since this discovery was confirmed by several independent observations (see \cite{Riess:1998cb,Perlmutter:1998np,deBernardis:2000sbo,Hanany:2000qf,Riess:2004nr,Spergel:2003cb,Spergel:2006hy} for a detailed discussion of the recent astronomical observations).
In order to explain the accelerated expansion, there exist two different approaches to solve the debate. One is the existence of mysterious dark energy (DE) and its possible extensions \i.e. modified theories of gravity (MTG). The idea of DE which has negative pressure occupied approximately 70\% of the energy density of our universe exists in a non-matter form. The simplest example of the DE is the cosmological constant $\Lambda$, as representing a constant energy density of the vacuum which satisfy the cosmological observations \cite{Zeldovich}. But it is plagued by a severe energy scale problem if it originates from the vacuum energy appearing in particle physics \cite{Weinberg}. This problematic nature of cosmological constant has motivated intense research for alternative theories of gravity extending the Einstein's theory of gravity.
This motivates the search for an alternative gravity theory that can address the present accelerating phase of the universe. It has been shown that alternative possibilities could give an adequate description of cosmological observations \cite{Jain:2010ka}. One of the simplest possible modification is the $f(R)$-gravity \cite{DeFelice:2010aj,Capozziello},
has attracted serious attention possibly because of its (deceptive) simplicity. A viable $f(R)$ gravity, where $f(R)$ is a generic function of the Ricci scalar $R$. This theory comes into the game as a straightforward extension of GR and to discuss a unified picture of both inflation and the accelerated expansion in more scientific ways \cite{Santos:2010tw,Carloni:2010ph,Dunsby:2010wg,Ketov:2010qz,Sotiriou:2008rp}. { In Ref. \cite{Astashenoka2015}, authors have studied a nonperturbative model of strange spherical objects in $f(R)=R+2\,\alpha\,R^2$ gravity theory, where $\alpha$ is a constant. They showed that the mass of the spherical objects increases when the value of the parameter $\alpha$ increases progressively. In addition,  Capozziello \textit{et al} \cite{111,222} have argued that the mass-radius profile undergoes modifications in $f(R)$ domain due to the presence of high order curvature terms such as $R^{2}$, $R^{3}$ etc.
For reviews of $f(R)$ theory in more details see Refs.
\cite{r4,r5,r8,r10,r11,r16,r18,r20,r21}. In addition to this, the first $f(R)$ work with approximately viable $f(R)$ to describe dark energy and inflation was proposed by Nojiri and his collaborators \cite{Nojiri1,Nojiri2,Nojiri3}. In this framework Hydrostatic equilibrium and stellar structure, stable neutron star models and extreme neutron stars have been discussed in details \cite{ Astashenok2014, Astashenok2013, Capozziello2011}. } 

In the same way, another alternative theory of gravity has been introduced, the so-called $f(R,T)$-gravity \cite{Harko:2011kv}.
This theory is based on the assumption of the gravitational field couples to the trace $T$ of the energy-momentum tensor of the matter. An interesting aspect of $f(R,T)$ theory is that it may provide an effective classical description of the quantum properties of gravity. Apart from a better understanding at the fundamental level, some results have been obtained with this theory. In an argument Houndjo \cite{Houndjo:2011tu} discussed transition of matter-dominated era to an accelerated phase
by assuming a special form of function $f(R,T)$ = $f_1(R)$ + $f_2(T )$.  The study of cosmological
solutions of $f(R, T)$ gravity was performed through the phase space analysis \cite{Shabani:2013mya}.
The other motivation is related to reconstructing $f(R,T)$ gravity from holographic dark energy; see, e.g.
 \cite{Houndjo:2011fb}. Other issues as, for example, cosmological and solar System Consequences \cite{Shabani1:2014zza}, anisotropic cosmology \cite{Sahoo:2017poz,Fayaz:2014bja},
 non-equilibrium picture of thermodynamics \cite{Sharif:2012zzd}, a wormhole solution \cite{Moraes:2017mir,Sahoo:2017ual}, and some other relevant aspects \cite{Saha:2014mua,Momeni:2015gka,Noureen:2015nja,Yousaf:2016lls}. 

Therefore, it is not possible to confirm or to disprove such theories based on the results of cosmology and compare them with the observational data.
{However, to establish a satisfactory gravity theory, it is important to study on the astrophysical level, e.g. using the relativistic stars. Some arguments for these theories come from the assumption that relativistic stars in the strong gravitational field could discriminate standard gravity from its generalizations.} Considering the case of $f(R,T)$ gravity, a large number of works on the evolution of compact stars are available in different literature. In this framework, hydrostatic equilibrium configuration of neutron stars and strange stars have been studied \cite{Moraes:2015uxq}. The structure of compact stars in $f(R,T)$ gravity was investigated recently in refs. \cite{Zubair:2015gsb,Das:2016mxq,Deb:2017rhd,Deb:2018sgt,Maurya:2019sfm,Maurya:2019iup}, whereas gravitational vacuum condensate star (gravastars) solution has been obtained in \cite{Das:2017rhi}.

Recently, Hansraj and Banerjee \cite{Hansraj:2018jzb} have studied stellar models within the context of $f(R,T)$ gravity, and showed that in some situation these theories
displays more pleasing behavior than its Einstein counterpart.
Motivated by these good antecedents, we extend Buchdahl's \cite{Buchdahl:1959rhi} spacetime from GR framework to the $f(R,T)$ gravity arena. The Buchdahl \textit{ansatz} is well-known solutions in GR with a clear geometric characterization of the associated spacetime metric will be prescribed to determine
the other. In particular, this \textit{ansatz} contains a wide range of models, each one is mathematically different from the other.  As a toy model, these solutions accurately describe the behavior of real astrophysical objects such as neutron stars, white dwarfs or strange star families \cite{Paul}. Thus, the information obtained from these celestial bodies has allowed a better understanding of the behavior of gravitational interaction in the strong field regime and some intricate processes of creation/annihilation of particles, among others.

From the aforementioned discussion one natural question arises, does the Buchdahl model carry over to $f(R,T)$ theory, which maintains the hydrostatic equilibrium and compatible with GR solution? To address this question, we would like to pass some remarks: I) We consider perfect
fluid matter distribution as of GR case. Because, our interest is to see the effects of $f(R,T)$ theory on compact stars model. In addition, we would like to compare our results with there similes in GR; II) Despite its simplicity, the chosen $f(R,T)$ model could be seen as Einstein gravity plus an effective cosmological constant; III) The Lagrangian  matter density $\mathcal{L}_{m}$ can be taken as $-p$ \i.e. the isotropic pressure. These are the main motivation for the extension of the Buchdahl model to more complicated
and general situations. An interesting aspect of our solution is that we use \emph{Gupta-Jasim} two steps methodology
for solving the differential equations. Further, we explore the obtained solution by studying all fundamental properties that any well behaved stellar structure should satisfied. These constraints include:

 \begin{itemize}
    \item{(a)} Positivity and finiteness of pressure and energy density
    everywhere in the interior of the star including the origin and
    boundary:
    \begin{eqnarray}
    0< p< \infty, ~~~~~ 0< \rho < \infty \nonumber \end{eqnarray}
    
    \item{(b)} Inside the fluid sphere the pressure and density should be monotonic
    decreasing functions with increasing radius. The pressure vanishes at the boundary $r=R$:
    \begin{eqnarray}
    \frac{dp}{dr}\leq0,~~~~~~\frac{d\rho}{dr}\leq0, ~~~~~~p(R)=0 \nonumber
    \end{eqnarray}
    
    \item{(c)} At the boundary of the star the interior solution should be matched with
    the Schwarzschild exterior solution, i.e. $ds^2_{-}=ds^2_{+}$. If follows
        \begin{eqnarray}
   e^{\nu(R)}=e^{-\lambda(R)}=1-\frac{2M}{R}. \nonumber
    \end{eqnarray}
   
    \item{(d)} Inside the fluid sphere the velocity of sound should everywhere be
    less than the speed of light
    \begin{eqnarray}
   0\leq v^2=\frac{dp}{d\rho}\leq 1. \nonumber
    \end{eqnarray}
 
      \item{(e)} The physical ways to characterize the energy conditions which are:
    \begin{itemize}
        \item Null energy condition: $\rho + p > 0$
        \item Weak energy condition: $\rho> 0$ \text{and} $\rho + p > 0$
        \item Strong energy condition: $\rho + 3p > 0$
        
        \item Dominant energy condition: $\rho \geq |p|$
    \end{itemize}
    
    should be satisfied.

    \item{(f)} The solution should be free from physical and geometric singularities i.e. $e^{\nu}$ and $e^{\lambda}$ in the range $0\leq r\leq R. $

     \item{(h)} It is the Buchdahl limit \cite{Buchdahl:1959rhi} for a perfect fluid sphere of radius $R$ and mass $M$, if $M/R \leq \frac{4}{9}$, then there is no equilibrium solution whatsoever.  The impact of this upper bound is that one cannot pack more matter into an object than the radius allows. As noticed in \cite{Goswami:2015dma}, this upper bound is larger than the Buchdahl-Bondi limit of GR in $f(R)$ gravity, whenever $f(R) \neq R$. Moreover,  Chakraborty and Sengupta \cite{Chakraborty:2017uku} established that an extra-massive stable star can exist in the context of Kalb-Ramond field in a four dimensional spacetime.
\end{itemize}

{Finally, it is worth mentioning that,  Buchdahl's solution has an enormous backrest as viable and tractable model describing compact configurations. In this respect, Vaidya and Tikekar \cite{Vaidya:1982vt} particularized the model giving a geometric meaning, prescribing specific 3-spheroidal geometries ($t$= \text{const.}) to 4-dimensional hypersurfaces. Note that this spheroidal condition  has been found very useful for finding exact analytic solutions of the Einstein field equations (EFEs) in GR and have important applications ranging from singularity free interior solutions to the physical understanding of relativistic phenomena \cite{Kumar:2018rlo}. Furthermore, using this situation Kumar \textit{et al.} \cite{Kumar:2005kuy,Kumar:2018hgm} have found an exact solution to the EFEs with an anisotropic matter distribution and  admitting conformal motion \cite{Bhar:2014jta}. In the framework of anisotropic hypothesis ``Buchdahl model" have been tested against astrophysical compact stellar objects (for review, see \cite{Maurya:2018kxg}).}

This article is organized as follows: Starting
with a brief introduction in Sect.~$\ref{sec1}$, we make a  review of the original $f(R,T)$ gravity in Sect.~$\ref{sec2}$ and present the formalism that allows us to construct stellar structure for spherically symmetric solutions in Sect.~$\ref{sec3}$. In Sect.~$\ref{sec4}$, the compact star models in frame of modified gravity with $f(R,T)= R+2\chi T$, are investigated in detail. For compact stars we use a well-known metric ansatz proposed by Buchdahl and find its solutions. In continuation with this we derive the field equations by using \textit{Gupta-Jasim} two steps method and studied solutions for positive and negative values of Buchdahl parameter $K$. We show our results and discussions in the same section and draw the final conclusions in Sect.~$\ref{sec5}$.



\section{$f(R, T$) gravity theory}\label{sec2}

In this section, we concisely review the viable modified theory of gravity,
as in the case of $f(R, T)$ gravity with $T$ being the trace of the stress-energy tensor, $T_{\mu\nu}$. The full action is
\begin{eqnarray}\label{1.1}
S=\frac{1}{16\pi}\int f(R,T)\sqrt{-g} d^{4}x+\int \mathcal{L}_{m}\sqrt{-g}d^{4}x, \label{eq2}
\end{eqnarray}
where $f(R,T)$ is the generic function of Ricci scalar $R$ with $g$ is the determinant of the metric tensor $g_{\mu\nu}$. We define the matter Lagrangian density,
related to the energy-momentum tensor as
\begin{eqnarray}
T_{\mu\nu}=-\frac{2}{\sqrt{-g}}\,\frac{\delta(\sqrt{-g}\,\mathcal{L}_m)}{\delta g^{\mu\nu}}, \label{eq3}
\end{eqnarray}
with the trace $T=g^{\mu\nu}T_{\mu\nu}$. Following the Ref \cite{Harko:2011kv}, we consider the case of Lagrangian density $\mathcal{L}_m$ of matter depends only on the metric tensor components $g_{\mu\nu}$.
Contracting Eq. (\ref{eq3}) gives
\begin{eqnarray}
T_{\mu\nu}=g_{\mu\nu} \mathcal{L}_m-\frac{2\,\partial(\mathcal{L}_m)}{\partial g^{\mu\nu}}. \label{eq4}
\end{eqnarray}
{
By varying the action (\ref{eq2}) of the gravitational field with respect to the metric tensor components $g^{\mu\nu}$ provides the following relationship:
\begin{multline}\label{ac}
\delta S =\frac{1}{16\pi} \int \Big[ f_{R}(R,T)R_{\mu\nu}\delta  g^{\mu\nu} + f_{R}(R,T)g_{\mu\nu}\Box \delta  g^{\mu\nu}
\\  
 -f_{R}(R,T) \nabla_{\mu}\nabla_{\nu} \delta g^{\mu\nu}+f_{T}(R,T)\frac{\delta(g^{\alpha\beta}T_{\alpha\beta})}{\delta g^{\mu\nu}}\delta g^{\mu\nu} \\   
 -\frac{1}{2}g_{\mu\nu}f(R,T)\delta g^{\mu\nu}+16 \pi
 \frac{1}{\sqrt{-g}}\frac{\delta(\sqrt{-g}\,\mathcal{L}_m)}{\delta g^{\mu\nu}} \Big]\sqrt{-g}d^{4}x ,
\end{multline}
where $f_R (R,T)={\partial f(R,T)}/{\partial R}$ and $f_T (R,T)={\partial f(R,T)}/{\partial T}$. According to Ref \cite{Harko:2011kv}, the variation of $T$ with respect to the metric tensor as
\begin{eqnarray}
\frac{\delta(g^{\alpha\beta}T_{\alpha\beta})}{\delta g^{\mu\nu}}\delta g^{\mu\nu}=T_{\mu\nu}+\Theta_{\mu\nu}.
\end{eqnarray}
The $\nabla_\mu$ denotes covariant derivative which is associated with the Levi-Civita connection of metric tensor $g_{\mu\nu}$ and box operator $\Box$ is defined by
\begin{eqnarray}
\Box\equiv\partial_\mu(\sqrt{-g}g^{\mu\nu}\partial_\nu)/\sqrt{-g}, ~~~~ \textrm{and} ~~~~ \Theta_{\mu\nu}=g^{\alpha\beta}\delta T_{\alpha\beta}/ \delta g^{\mu\nu}. \nonumber
\end{eqnarray}
Now, partially integrating the second and third terms in Eq. (\ref{ac}), one can obtain the field equations of the $f (R, T)$ gravity model as
\begin{eqnarray}\label{1.2}
\left( R_{\mu\nu}- \nabla_{\mu} \nabla_{\nu} \right)f_R (R,T) +\Box f_R (R,T)g_{\mu\nu} - \frac{1}{2} f(R,T)g_{\mu\nu}  \nonumber \\ = 8\pi\,T_{\mu\nu}  - f_T (R,T)\, \left(T_{\mu\nu}  +\Theta_{\mu\nu}\right).
\end{eqnarray}
We see from this equation that when $f (R, T)$ $\equiv f(R)$, the above equation reduces to $f(R)$ gravity field equations. In particular, if $f (R, T)$ $\equiv R$ then standard Einstein's field equations are recovered in GR.}

To reach the expression of the covariant derivative of the energy-momentum tensor and extract the one of
the algebraic function, we perform the covariant derivative of Eq.(\ref{1.2}), as \cite{Alvarenga:2013syu}
\begin{eqnarray}\label{1.3}
\nabla^{\mu}T_{\mu\nu}=\frac{f_T(R, T)}{8\pi -f_T(R,T)}\bigg[(T_{\mu\nu}+\Theta_{\mu\nu})\nabla^{\mu}\ln f_T(R,T)\nonumber\\ +\nabla^{\mu}\Theta_{\mu\nu}-\frac{1}{2}g_{\mu\nu}\nabla^{\mu}T\bigg].~~~~~~
\end{eqnarray}

It is straightforward to see that the stress-energy momentum tensor $T_{\mu\nu}$ in $f(R,T)$ gravity is not conserved as a view point of Einstein general relativity (GR) due to presence of nonminimal matter-geometry coupling in the formulation.
By using Eq. (\ref{eq4}), the tensor $\Theta_{\mu\nu}$ is defined as
\begin{equation}\label{1.4}
\Theta_{\mu\nu}= - 2 T_{\mu\nu} +g_{\mu\nu}\mathcal{L}_m - 2g^{\alpha\beta}\,\frac{\partial^2 \mathcal{L}_m}{\partial g^{\mu\nu}\,\partial g^{\alpha\beta}}.~~~
\end{equation}

Henceforth, in order to facilitate a direct comparison with the work of Buchdahal \cite{Buchdahl:1959rhi}, we follow his conventions.
For stellar configurations, one can assume a spherically symmetric metric with coordinates ($t$, $r, \theta, \phi$)
in the following form
\begin{equation}\label{eq1}
ds^{2} = e^{\nu(r) } \, dt^{2}-e^{\lambda(r)} dr^{2}-r^{2}(d\theta ^{2} +\sin ^{2} \theta \, d\phi ^{2}),
\end{equation}
where $\nu (r)$ and $\lambda(r)$ are arbitrary functions of the radial coordinate $r$ only.  We use the natural system of units with $G$=$c$=1. Further, we assume that the interior of the star is filled with a perfect fluid source, and the stress tensor is given by
\begin{equation}\label{1.51}
T_{\mu\nu}=(\rho+p) u_\mu u_\nu-p g_{\mu\nu},
\end{equation}
where ${u_{\nu}}$ is the four velocity, satisfying $u_{\mu}u^{\mu}= 1$ and $u^{\mu}\nabla_{\nu}u_{\mu}=0$.
Here, $\rho$ is the matter density and $p$ is the isotropic pressure.  Since, we choose a further  assumption, namely, $\mathcal{L}_m= -p$, according to the definition suggested in \cite{Harko:2011kv}, the tensor (\ref{1.4}) yields
\begin{eqnarray} \label{1.5}
\Theta_{\mu\nu}=-2T_{\mu\nu}-p\,g_{\mu\nu}.
\end{eqnarray}

We then assume a more plausible and simple model, namely, $f(R,T)= R+2\chi T$, the usual Einstein-Hilbert term plus a $T$ dependent function $f(T )$ \cite{Harko:2011kv}. This interesting model has some significant advantages, for example, this model can describe inflationary era as well as an accelerated expansion phase. In this framework we obtain singularity free spacetime. Thus, using the linear expression and Eq. (\ref{1.2}), the Einstein tensor reduce to  
\begin{eqnarray}\label{1.52}
G_{\mu\nu}= 8\pi\,T_{\mu\nu}+\chi T g_{\mu\nu}+2\chi(T_{\mu\nu}+p\,g_{\mu\nu})=8\pi \tilde{T}_{\mu\nu}.~~~
\end{eqnarray}

Note that field equation (\ref{1.2}) reduce to Einstein field equations when $f(R,T)$ $\equiv R$.{ Studying such particular linear assumption ($f(R,T)= R+2\chi T$) has widely accepted to address cosmological as well as
astrophysical solutions (see introduction).} By substituting the value of $f(R,T)=R+2\chi T$ in Eq. (\ref{1.3}), we obtain
\begin{eqnarray}\label{1.6a}
\left(8\pi+ 2\chi\right)\nabla^{\mu}T_{\mu\nu}=-2\chi\big[g_{\mu\nu}\nabla^{\mu}T+2\,\nabla^{\mu}(p\, g_{\mu\nu})\big].
\end{eqnarray}
In that follows, if $\chi \rightarrow 0$ one can recover the conservation principle of energy-momentum which plays a crucial role in Einstein's gravitational theory. Nevertheless, in the next section we focus our attention on a stellar model that contain  perfect fluid i.e. the flow of matter is adiabatic, no heat flow, radiation, or viscosity is present.


\section{Einstein field equations}\label{sec3}

Let us investigate the non-zero components of the field equations for spherically symmetric line  element (\ref{eq1}), which are \cite{Hansraj:2018jzb}
\begin{eqnarray}
\tilde{\rho} &=&\frac{e^{-\lambda}}{8\pi}\left(-\frac{1}{r^2}+\frac{\lambda^{\prime}}{r}+\frac{e^{\lambda}}{r^2}\right),\label{deff} \\ ~~~
\tilde{p}&=&\frac{e^{-\lambda}}{8\pi}\left(\frac{1}{r^2}+\frac{\nu^{\prime}}{r}-\frac{e^{\lambda}}{r^2}\right),\label{preff} \\
\tilde{p}&=&\frac{e^{-\lambda}}{32\pi}\left(2\nu^{\prime\prime}+\nu^{\prime2}-\lambda^{\prime}\nu^{\prime}+2\frac{\nu^{\prime}-\lambda^{\prime}}{r}\right).\label{pteff}~~~
\end{eqnarray}
with,
\begin{eqnarray}
\tilde{\rho}=\rho+\frac{\chi}{8\,\pi}(3\rho-p),\nonumber\\
\tilde{p}= p -\frac{\chi}{8\,\pi}(\rho-3 p), \nonumber
\end{eqnarray}
where the prime denotes differentiation with respect to the radial coordinate $r$. Using Eqs. (\ref{deff}-\ref{pteff}), one can obtain another additional equation which is
\begin{eqnarray}
\frac{\nu^{\prime}}{2}\,(p+\rho)+\frac{dp}{dr}=\frac{\chi}{(8\pi+2\chi)}\,(p^{\prime}-\rho^{\prime}).\label{TOV}
\end{eqnarray}
Note that Eq. (\ref{TOV}) reduces to the hydrostatic equilibrium condition of general relativity
when $\chi=0$. Now, using Eqs. (\ref{deff}) and (\ref{preff}), we rewrite the modified equations
in terms of pressure ($p$) and energy density ($\rho$), which are
\begin{eqnarray}
\rho&=&\frac{\pi\,}{(8\pi^2+6\pi\chi+\chi^2)}\,[ (8\pi+3\chi)\,\tilde{\rho}\,+\chi\,\tilde{p}\,],\label{rhof}\\
p&=&\frac{\pi\,}{(8\pi^2+6\pi\chi+\chi^2)}\,[ (8\pi+3\chi)\,\tilde{p}\,+\chi\,\tilde{\rho}\,].\label{pf}~~
\end{eqnarray}
where $\chi\ne -2\pi~ \& -4\pi$. While the equation of pressure isotropy $G^1_1=G^2_2$ reduces to
\begin{eqnarray}
r^2\,(2\nu^{\prime\prime}+\nu^{\prime 2} -\nu^{\prime}\lambda^{\prime})-2r(\nu^{\prime}+\lambda^{\prime})+4 (e^{\lambda}-1) =0.~~~~\label{iso}
\end{eqnarray}
Interestingly the isotropy equation for $f(R,T)$ gravity is the same for the ordinary Einstein's equations with a perfect fluid source. The mass of the star is now due to the total contribution of the energy density of the matter and that can be determined by the Eq. (\ref{deff}). The mass takes the new form as
\begin{eqnarray}
m=4\pi\int_0^r{\rho\, r^2\,dr}= \frac{4\,\pi^2\,(8\pi+3\chi)}{(8\pi^2+6\pi\chi+\chi^2)}\int_0^r{\tilde{\rho}\, r^2\,dr}\nonumber\\+\frac{4\,\pi^2\,\chi}{(8\pi^2+6\pi\chi+\chi^2)}\int_0^r{\tilde{p}\,r^2\,dr},\label{eq19} ~~
\end{eqnarray}
In this regard, constant parameter $\chi$ plays an important role for determining the stellar structure. Classically, the $f(R,T)$  gravity recovers the same physics as the general relativity with $\chi$= 0. Observing the Eq. (\ref{iso}), which serves as the master differential equation in this analysis contains two necessary constants of integration. However, one can accommodate the constants in terms of the mass $M$ and radius $R$ of the distribution by solving the linear matching equations.

As usual, all astrophysical objects are immersed in vacuum spacetime and at the juncture interface we match the interior spacetime to an appropriate exterior vacuum region. Here, we are working with uncharged matter distributions and then outer spacetime is described by Schwarzschild solution,
\begin{eqnarray}
\label{eq19a}
ds^{2}= \left(1-\frac{2M}{r}\right)dt^{2}-\frac{dr^{2}}
{1-\frac{2M}{r}}-r^{2}(d\theta^{2}+\sin^{2}\theta d\phi^{2}),
\end{eqnarray}
{where $M$ is the Schwarzschild mass which concides with total mass within the fluid sphere at the surface of the object defined by $\Sigma=r=R$. So, the first fundamental form demands the continuity of the metric potentials across the boundary. Explicitly it reads}
\begin{equation}
    e^{-\lambda(R)}=e^{\nu(R)} = 1-\frac{2M}{R}.\label{eq21}
\end{equation}
{On the other hand the second fundamental form dictates 
\begin{eqnarray}
p(R)=0,    
\end{eqnarray}
this condition says that object can not expand indefinitely, in consequence this constraint determines the object size \i.e, its radius.}
In next, we will investigate stellar structure with Buchdahl assumption as a  metric potential in the $f(R,T)$ theory to solve the field
equations.

\section{The Buchdahal Solutions}\label{sec4}

The Buchdahl solution was generated through a mathematical ansatz on the static, spherically symmetric fluid spheres of Einstein's equations by Buchdahl \cite{Buchdahl:1959rhi},
which cover almost all physically tenable known models. The widely studied metric ansatz given by
\cite{Maurya:2018kxg}
 \begin{equation}
 e^{\lambda}=\frac{K\,(1+Cr^2)}{K+Cr^2}, ~~~~~ \textrm{when} ~~~ K < 0~~~ \textrm{and}~~ K > 1, \label{lambda1}
\end{equation}
where $K$ and $C$ are two parameters that characterize the geometry of the star.  Here, we extend the following work \cite{Maurya:2018kxg}, which was devoted in  describing a class of relativistic stellar solutions for generalized Buchdahl dimensionless parameter $K$.

An interesting aspect of the Buchdahl solution is that one can recover the interior Schwarzschild solution when $K=0$ and for $K=1$ the hypersurfaces $\{t = \mbox{constant} \}$ are flat. Further, if we assume $C = -K/R^2$, one can recover the Vaidya and Tikekar \cite{Vaidya:1982vt} solution and for $K=-2$ we recover Durgapal and Bannerji \cite{Durgapal} solution. Several authors \cite{Mkenyeleye:2014dwa,Gupta:2003guj,Kumar:2005kuy} subsequently proved that it is a viable physical model, and showed that it could be used to classify some of the previously known exact solutions.

In order to solve the field equations (\ref{deff}-\ref{pteff}), we introduce the transformation $e^{\nu}=\Psi^2$ and $\xi=\sqrt{\frac{K+Cr^2}{K-1}}$ \cite{Kumar:2018rlo,Kumar:2018hgm}. For more clear sighted, we consider the Gupta-Jasim \cite{Gupta:2003guj} (see Refs. \cite{Maurya:2018kxg} for complete discussion)
two step method for solving the system of equations. Starting with the pressure isotropic Eq. (\ref{iso}), and using Eq. (\ref{lambda1}) the corresponding differential equation will then
\begin{eqnarray}
(1-\xi^2)\,\frac{d^2\Psi}{d\xi^2}+\xi\,\frac{d\Psi}{d\xi}+(1-K)\Psi=0, \label{hyper}
\end{eqnarray}
In this framework we consider mainly two cases  for the spheroidal parameter $K$ with $C>0$.

\subsection{{Case I. For $K < 0$~~ i.e  K is negative}}

Now, differentiate the Eq. (\ref{hyper}) with respect to $\xi$ and use another  substitution $\xi=\sin{x}$ and $\frac{d\Psi}{d\xi}=Y$,  we have
\begin{eqnarray}
\frac{d^2Y}{dx^2}+(2-K)Y=0,~~~~\textrm{for}~K<0,~~\xi=\sin x, \label{eq22}
\end{eqnarray}
where $\frac{dY}{dx}=\cos{x}~\frac{d^2\psi}{d\xi^2}$ and $\frac{d^2Y}{dx^2}=\cos^2{x}~\frac{d^3\psi}{d\xi^3}-\sin{x}~\frac{d^2\psi}{d\xi^2}$, respectively.  The above Eq. (\ref{eq22}) is a homogeneous differential equation of second order with constant coefficients. In this case Eq. (\ref{eq22}) leads to
\begin{eqnarray}
 Y(x)=a_{1} \sin (n\,x)+b_{1} \cos (n\,x),~\textrm{if}~~2-K=n^2, ~K<0.~\label{11}
\end{eqnarray}
where $a_{1}$ and $b_{1}$ are arbitrary constants of integration.
 To obtain the complete solution we re-substitute  $\frac{d\Psi}{d\xi}=Y$ and $\frac{d^2\Psi}{d\xi^2}=\frac{dY}{dx}\frac{dx}{d\xi}$ into hypergeometric Eq. (\ref{hyper}), we get $\Psi$ as

\begin{eqnarray}\label{eq24}
\Psi(x)=\frac{\sin x [b_{1}+ a_{1}\tan(n\,x)]- n \cos x [b_{1} \tan(n\,x)-a_{1} ]}{\textrm{sec}(n\, x)(1 - n^2)},
\end{eqnarray}
where $n=\sqrt{2-K}$ and $K<0$.

We calculate in detail the components of Eqs. (\ref{deff}-\ref{preff}) using the Eqs. (\ref{lambda1}) and (\ref{eq24}), and we find
\begin{eqnarray}
&&\rho_E = \frac{C\,[3-K+(K-1)\,\sin^2x]}{8\,\pi\,K\,(K-1)\,\cos^4x} ,~~\label{eq25}\\
&&p_E = \frac{C}{8\,\pi\,K\,\cos^2x}\bigg[\frac{2\,\sin x\,[a_1 \tan (n\,x)+b_1 ]}{\,(1-K)\,\Psi(x)\,\textrm{sec}(n\,x)}+1\bigg],\label{eq261}~~~~~~
\end{eqnarray}
where $\rho_E$ and $p_E$ are the Einstein energy density and pressure, respectively. When the Einstein metric components are inserted into the $f(R, T)$ counterparts, we obtain
\end{multicols}
\begin{eqnarray}
  &&\rho_f = \frac{C}{(8\pi^2+6\pi\chi+\chi^2)}\bigg[\frac{(8\pi+3\chi)\,[3 - K + (K-1)\sin^2x]}{8 K\,(K-1)\,\cos^4x}+ \frac{\chi\,[2\,\sin x\,[ a_1 \sin (n\,x)+b_1\,\cos(n\,x) ]+(1-K)\,\Psi(x)]}{8 K\,(1-K)\,\Psi(x)\,\cos^2 x}\bigg], ~~~~~~~~\label{eq27}\\
&&p_f = \frac{C}{(8\pi^2+6\pi\chi+\chi^2)}\bigg[\frac{\chi\,[3 - K + (K-1)\sin^2x]}{8 K\,(K-1)\,\cos^4x} + \frac{(8 \pi+3\chi)\,[2\,\sin x\,[ a_1 \sin (n\,x)+b_1\,\cos(n\,x) ]+(1-K)\,\Psi(x)]}{8 K\,(1-K)\,\Psi(x)\,\cos^2 x}\bigg]. ~~~~~~~~~\label{eq28}
\end{eqnarray}
\begin{figure*}[!htp]
   \centering
   \includegraphics[width =7cm]{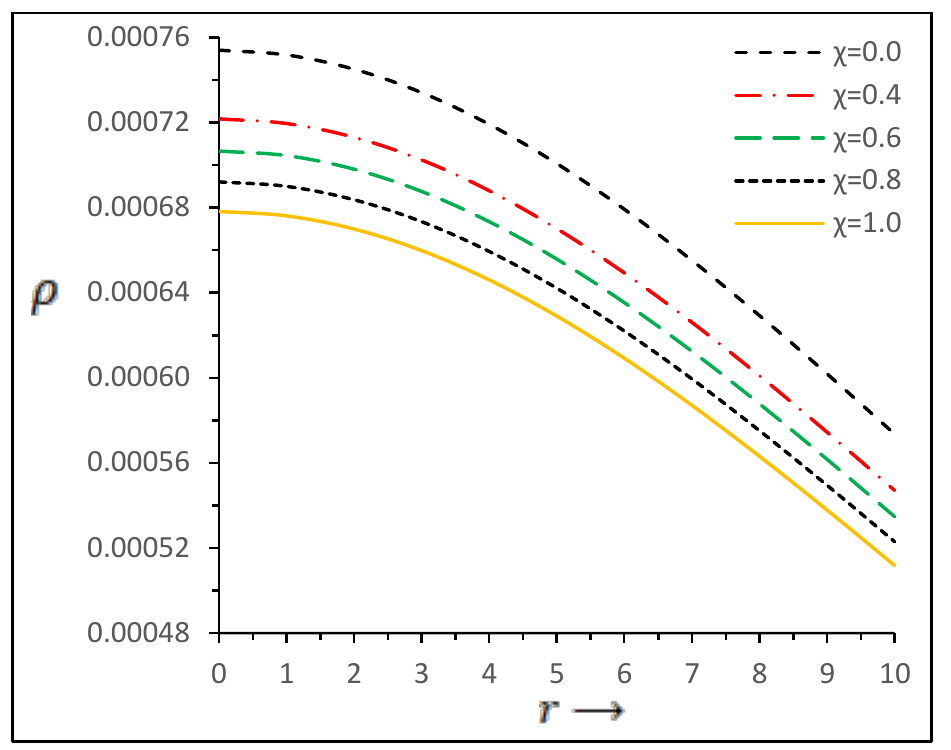}~~~ \includegraphics[width=7cm]{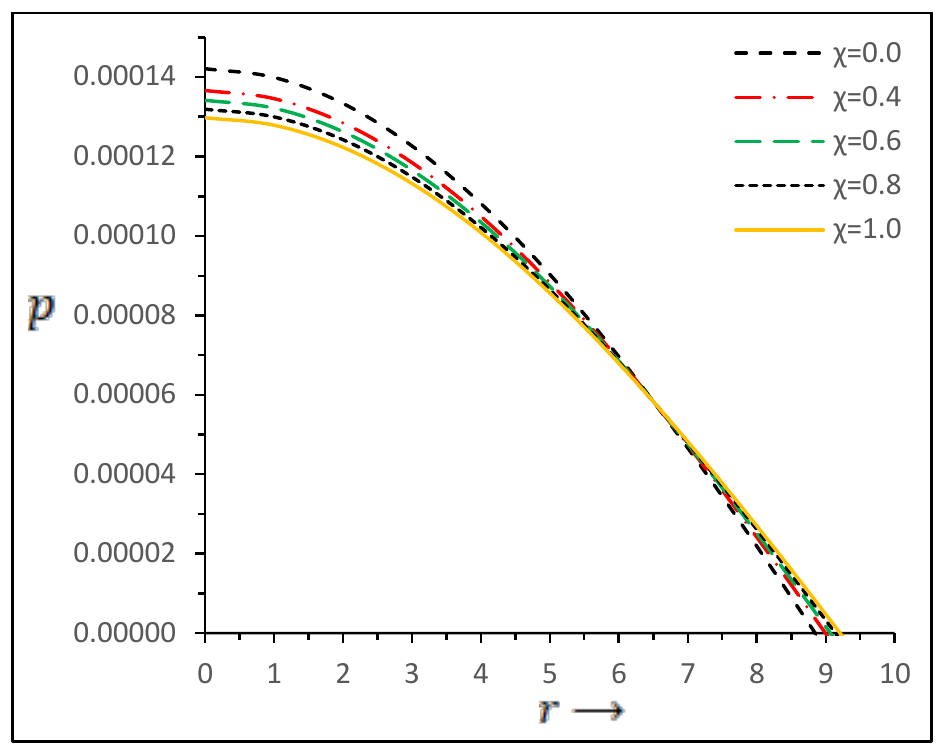}
   \caption{ \label{fig1}The energy density and pressure against radial coordinate $r$ have been plotted for modified gravity model $f(R,T)$ and in GR for comparison for some different values $\chi$ and fixed negative value of $K= -0.4$. For plotting we consider $\chi =0$ ( dashed black curve for GR case) reveals that for a mass of $M=1.3 M_{\odot}$, the radius goes upto $R=8.849$ Km, whereas $\chi =0.4, 0.6, 0.8$ and 1 the radius goes as high as $8.997$, $9.069$, $9.141$ and $9.211$, respectively. In all cases, it is considered the constant $C = 1.8047\times 10^{-3}km^{-2}$. }
\end{figure*}
\begin{figure*}[!htp]
   \centering
   {\includegraphics[width=7.5cm]{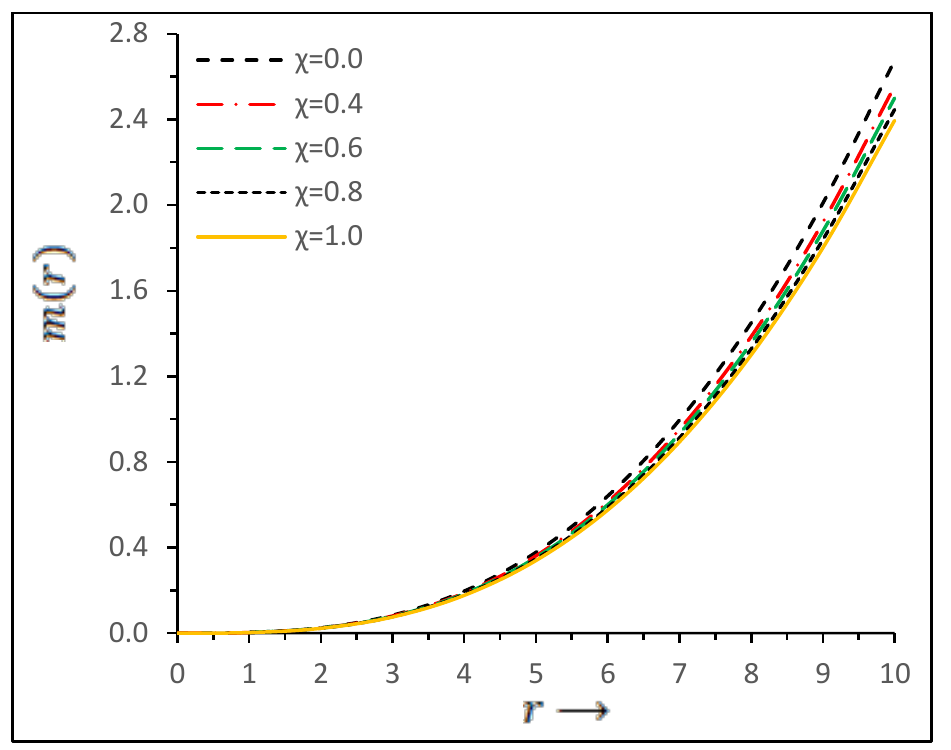}~~~~ \includegraphics[width=7.5cm]{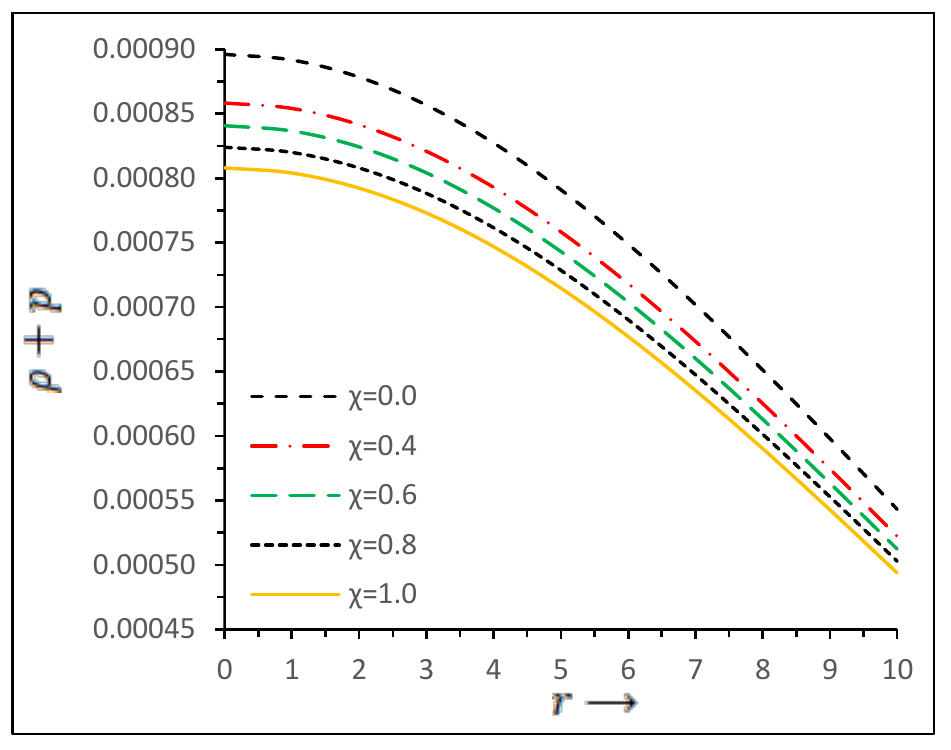}
\caption{\label{fig2} Gravitational mass (left diagram) and energy condition  (right diagram) versus radial coordinate $r$ have been plotted. For NEC is determined by the condition $\rho+p >0$ (see introduction). We have employed the same data set as used in Fig. \ref{fig1}.}} 
\end{figure*}
\begin{figure*}[!htp]
   \centering
   {\includegraphics[width=7.5cm]{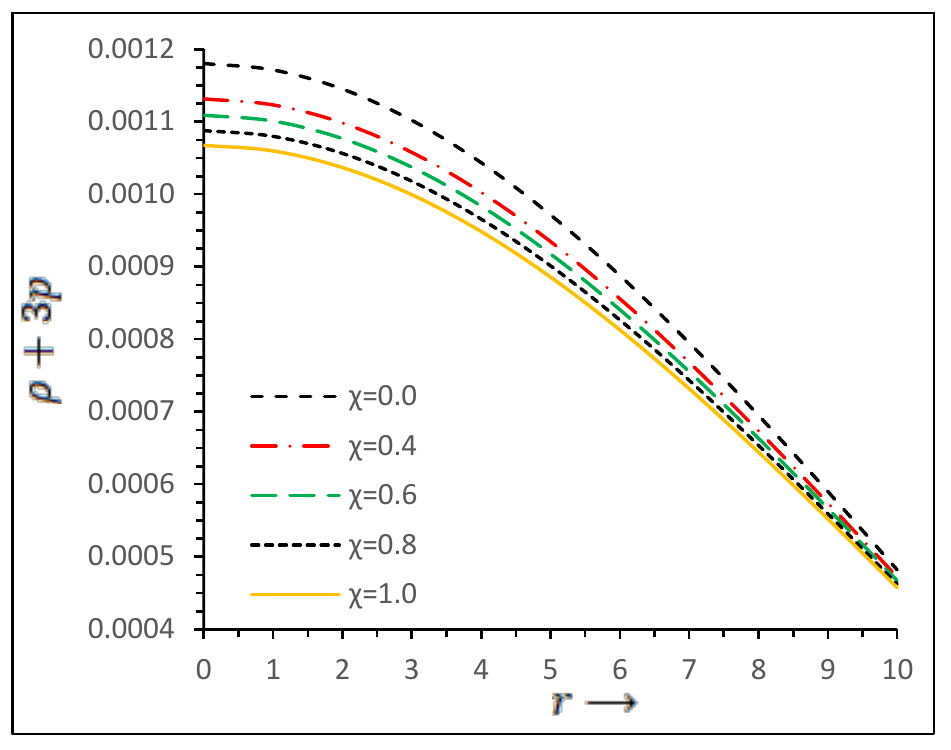}~~~~~ \includegraphics[width=7.5cm]{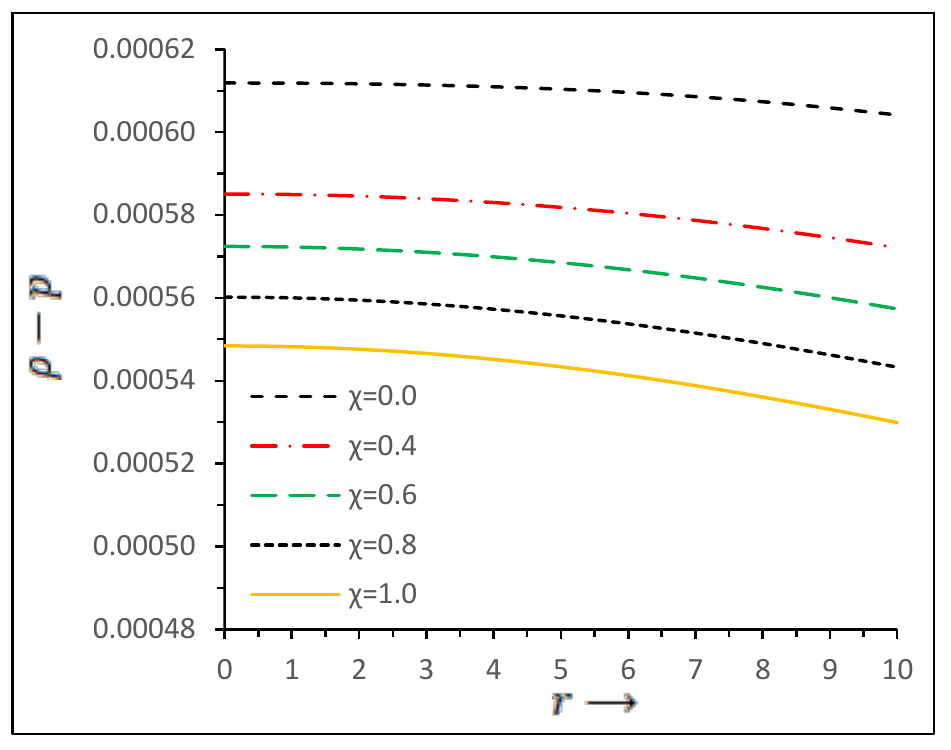}
\caption{\label{fig3} The SEC and DEC against the radial coordinate $r$ have been plotted on the left and right panels
for some values of $\chi$. Of course, the physical form of the WEC is straightforward:  total energy density of matter field $\rho >0$ and satisfied  $\rho +p>0$.}}
\end{figure*}
\begin{multicols}{2}
Now, we study for relativistic compact stars depending on two parameters $K$ and $\chi$, and keeping the other
parameters same for $f(R, T)$ gravity and as of GR (when $\chi$ $\rightarrow 0$).  The behavior of the energy density $\rho$, the radial pressure $p$ and the mass $ m/M_{\odot}$ as a function of the radial coordinate are presented in Fig.~(\ref{fig1}-\ref{fig3}) for different values of $\chi$. Finally, we move on to describe the results obtained from our calculations, which are illustrated in Figs.~(\ref{fig1}-\ref{fig2}), for the GR case ($\chi = 0$) and for the $f(R, T)$  gravity ( different values of $\chi$).  As one can see from Fig.~\ref{fig1} that the pressure (energy density)
inside the star is positive and monotonically decreasing function towards the boundary, and reaches the value zero on the star surface.

The mass-radius relation is represented in Fig.~\ref{fig2} (left panel).  Note that if the radius is increasing with the mass, the $M/R$ ratio is also increasing,  but much slower for the greater values of $\chi$.  But it is worth noticing that  the maximum mass attended from standard GR for $\chi=0$ as evident in Table \ref{table1}. As a result GR has profound effects on the critical mass of compact stars. For the case of modified gravity the effect caused by the term $2 \chi T$ which affects the structure of the star.
The stellar configurations have been analyzed  for several values of the important physical parameters in the tabular format (see Table \ref{table1}).
All properties of this class of stars have been obtained by solving the system of Eqs. (\ref{eq25}-\ref{eq28}) for the $\chi$ =0, 0.4, 0.6, 0.8 and 1, respectively with $K= -0.4$
after arduous fine tuning. Interestingly, a stellar mass of $M=1.3 M_{\odot}$, the pressure value zero on the star surface goes as high as $8.997$, $9.069$, $9.141$ and $9.211$, for increasing  the value of $\chi$ i.e. when the radius of star increasing. One can see from Table \ref{table1} that the deviation of the central density and pressure from GR, in principle, is higher than the $f(R, T)$ model. In the plots the dashed black curve for GR case, while the other curves represent the modified gravity throughout this work. {Finally, we report the results for the model obtained in \cite{Maurya:2018kxg}.
It can be said that normal GR conditions ($\chi$ = 0 and $K<0$ ) yield higher value of gravitational mass of strange/compact stars. It is also evident that 
massive neutron stars (NSs) configurations, consistent
with observational estimates from PSR J0348-0432, 4U 1538-52 and Her X-1 \textit{etc}, values of
$\chi$ and $K<0$ must be very low comparing to 
appreciable deviations from the linear model of $f(R,T)$-gravity.}

In order to go further we discuss energy conditions and physical quantities, respectively. In the case of energy conditions according to  classical field theories of gravitation, we have analyzed the properties of the  null energy (NEC), weak energy (WEC), strong energy (SEC) and dominant energy conditions (DEC),  respectively. It is interesting to point out that all conditions hold simultaneously in the framework of modified and classical gravity, as evident in  Fig.~(\ref{fig2}-\ref{fig3}). We have used the same parameters values  as indicating in Fig.~\ref{fig1}.

A more physical model should automatically account for sound speed for perfect fluid distribution. It is obvious that the velocity of sound is less than the velocity of light i.e. $0< v^2=dp/d\rho <1$. 
\end{multicols}
For our stellar model the speed of sound is given by (for both cases)
\begin{eqnarray}
&&\left(\frac{dp}{d\rho}\right)_E=\frac{(1 - K)\, \left[2\Omega_{f_ 2} (x) - a_1\,\big(\sin (2\,n \,x) v_{11}(x) + 2\,\Omega_{f_3}(x)\big)\right]\,\cot x\,\cos^4x\,}{[K-9 + (K-1)\,\cos (2 x)]\,\big[\sin (x) [b_1 \cos (n\,x) +a_1 \sin (n\,x)] + v_{12}(x)\big]^2},~~~~~~\\
&&\left(\frac{dp}{d\rho}\right)_f=\frac{(8\pi+3 \chi)\,\left[2\,\tan x\,v_{13}(x)-\Omega_ {f_ {11}} (x)\right] + 2 \chi \tan x - \chi\, \Omega_{f_{12}} (x) + 2\,\chi\tan x \Omega_{f_{14}}(x)}{-\chi\, \Omega_ {f_ {11}} (x) + (8\pi+3\chi)\,\left[2\,\tan x - \Omega_{f_{12}} (x) + 2\,\tan x\,\Omega_ {f_{14}} (x)\right] + 2\,\tan x\,\chi \,v_{13}(x)}.~~~~~~
\end{eqnarray}
\begin{figure*}[!htp]
\centering
     \includegraphics[width=7.5cm]{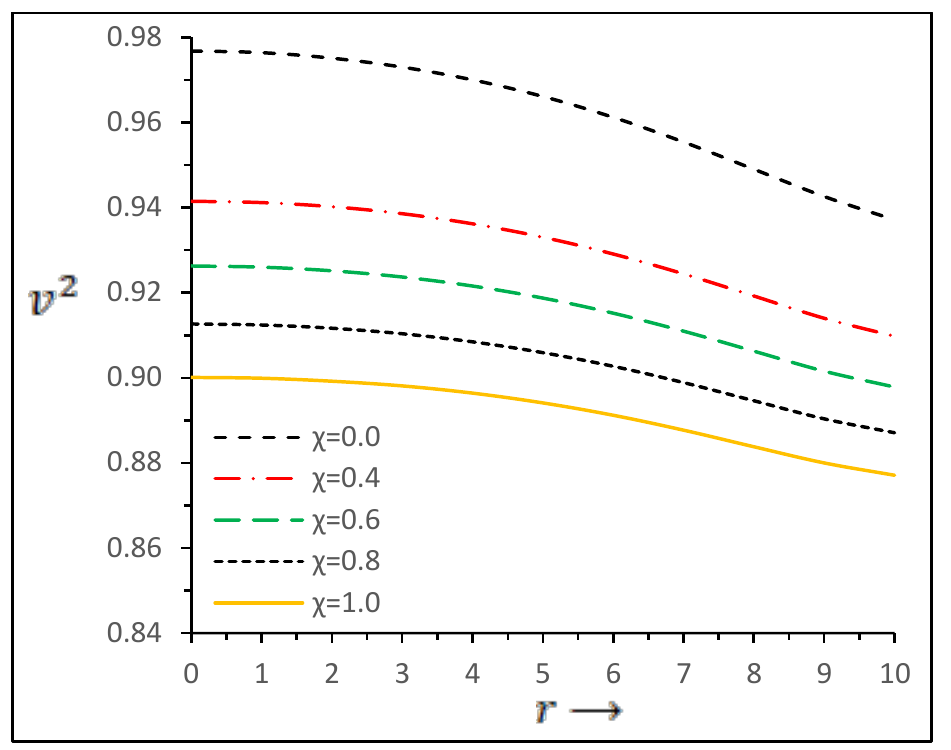}
\caption{\label{fig4}The velocity of sound versus radial coordinate $r$ have been plotted for $\chi =0$ (dashed black curve) $\chi =0.4$ (dash-dotted red curve), $\chi =0.6$ (long dashed green curve), $\chi =0.8$ (small dash black curve) and
$\chi =1$ (yellow curve).}
\end{figure*}
\begin{multicols}{2}
 In our analytical approach, we use graphical representation to represent the velocity of sound due to complexity of the expression. In this case, we clearly observe form Fig.~\ref{fig4} that velocity of sound is decreasing  away from the centre. Note that for GR case the sound speed is more closer to the velocity of light, but for this situation is more realistic for $f(R, T)$ gravity. The used coefficients in above equations are given in Appendix. 
\end{multicols}

{\centering{\subsection{{Case II. For $0<K <1 $~~ i.e.  K is positive}}}}
Here, we will report the solution for $0<K <1 $. As already mentioned that the energy density for Buchdahal model in GR is $\tilde{\rho}=\frac{4C(K-1) (3+Cr^2)}{32\pi(1+Cr^2)^2}$. Thus, it seems interesting
that within the limit of $0<K<1$, the energy density is negative due to the presence of $(K-1)$ term. Therefore, given the value of $K$, obtained solution is not physically valid in GR. Now, the density expression for $f(R,T)$ gravity could be determined from Eq. (\ref{rhof}), which involves the pressure term also. On the other hand, pressure term 
involves the metric potential $e^{\nu}$ and $e^{\lambda}$. However, in determining the metric potential $\nu$ by solving of hypergeometric Eq. (\ref{hyper}) we have to use the transformation $\xi=\sqrt{\frac{K+Cr^2}{K-1}}$. {It is worth noticing that the transformation is not valid under the square root of $0<K<1$, as there is no real solution. This implies that the pressure will not be physically valid. Therefore, such an analysis for GR and $f(R,T)$ gravity, however, is not physically valid.}\\

{\centering{\subsection{{Case III. For $K >1 $~~ i.e.  K is positive}}}}

In order to conduct further investigations, we extend our analysis for positive values of $K$. With the purpose of solving the Eq. (\ref{hyper}), we substitute $\frac{d\Psi}{d\xi}=Y$,  and $\xi=\cosh y$. Thus, we have
\begin{eqnarray}
\frac{d^2Y}{dy^2}+(K-2)Y=0,~~~~\textrm{for}~K>1,~~\xi=\cosh y.~~ \label{eq31}
\end{eqnarray}
This is a homogeneous differential equation of second order with constant coefficients. Now, we will classify the solution to the following cases

\begin{eqnarray}
{\textrm{Case IIIa:}} && Y(y)= a_{2}\,\cos (m\,y)+b_2\,\sin(m\,y),~~~~\textrm{if}~K-2=m^2~\textrm{and} ~~K>2.~~~~~~~~\label{33a}\\
{\textrm{Case IIIb:}}&& Y(y)= c_{2}\,\cosh(m\,y)+d_2\,\sinh(m\,y),~~~~\textrm{if}~2-K=m^2~\textrm{and}~1<K<2.~~~~~~~~~~~\label{33b}\\
{\textrm{Case IIIc:}}&&  Y(y)= e_{2}+f_{2}\,y,~~\textrm{if}~2-K=0.\label{33c}
\end{eqnarray}
where $a_{2}$, $b_{2}$, $c_2$, $d_2$, $e_2$ and $f_2$ are arbitrary constants of integration with $y=\cosh^{-1}\xi= \cosh^{-1}\sqrt{\frac{K+Cr^2}{K-1}}$. For simplification we substitute   $\frac{d\Psi}{d\xi}=Y$ and $\frac{d^2\Psi}{d\xi^2}=\frac{dY}{dy}\frac{dy}{d\xi}$ into hypergeometric Eq. (\ref{eq31}), and we get $\Psi$ as,

\begin{eqnarray}
{\textrm{Case IIIa}}:&&\hspace{-0.6cm}\Psi(y)=\frac{[ a_2 \cos(m\,y)+b_2\,\sin(m\,y) ]+m\,\tanh y\,[ a_2 \sin(m\,y)-b_2 \cos(m\,y)]}{(K-1)\, sech y} ~~~~~~\textrm{for}~~m=\sqrt{K-2}~\textrm{and}~K>2,~~~~\label{11b}\\
{\textrm{Case IIIb:}}&&\hspace{-0.6cm}\Psi(y)=\frac{m \tanh x\,[c_2 \sinh (m\,y)+d_2\cosh(m\,y)]-[d_2\sinh (m\,y)+c_2\cosh(m\,y)]}{(1-K)\,sech\, y}~\textrm{for}~m=\sqrt{2-K}~\textrm{and}~1<K<2,~~~~\label{12b}\\
{\textrm{Case IIIc:}}&&\hspace{-0.6cm} \Psi(y)= f_2\,[ y\,\cosh y-\sinh y ]+e_2\,\cosh y,~~~~~~\textrm{for}~~~K=2.
\label{12c}
\end{eqnarray}
where, $\Psi_{2}(x)=\cosh(m\,y)+d_2\,\sinh(m\,y)$. Recalling the Eqs. (\ref{deff}) and (\ref{preff}) and plugged into the relevant equation  we obtain the expression of energy density and pressure corresponding to Einstein and $f(R,T)$- gravity for three separate cases, as follows:\\

{\centering{\subsubsection{\textbf{Case IIIa: $K-2= m^2$,~~\textrm{and}~~$K>2$}}}}

Let us now  improve the above considerations by taking
into account $f(R, T)$ models with constant $\chi \neq 0$.  Thus, positive $K$-value provides a further set of expression for pressure and density, which are given by,
\begin{eqnarray}
&& \hspace{-0.95cm} \rho_E =\frac{C\,[ 3-K+(K-1)\,\cosh^2y ]}{8\,\pi\,K\,(K-1)\,\sinh^4y} ,~~\label{dE2a}\\
&& \hspace{-0.95cm} p_E =\frac{C}{8\,\pi\,K\,\sinh^2y}\bigg[\frac{2\,[a_2 \cos(m\,y)+b_2 \sin(m\,y)]}{(K-1)\,\Psi(y)\,sech y}-1\bigg] ,\label{pE2a}\\
&& \hspace{-0.95cm} \rho_f = \frac{C}{(8\pi^2+6\pi\chi+\chi^2)}\bigg[\frac{(8\pi+3\chi)[3-K+(K-1)\cosh^2y ]}{8\,K\,(K-1)\,\sinh^4y}+\frac{\chi}{8\,K\,\sinh^2 y}\bigg(\frac{2\,[a_2 \cos(m\,y)+b_2 \sin(m\,y)]}{(K-1)\,\Psi(y)\,\textrm{sech}\,y}-1\bigg)\bigg],~~\label{17a}\\
&& \hspace{-0.95cm}p_f = \frac{C}{(8\pi^2+6\pi\chi+\chi^2)}\bigg[\frac{\chi\,[ 3-K+(K-1)\,\cosh^2y ]}{8\,K\,(K-1)\,\sinh^4y}+ \frac{(8\pi+3\chi)}{8\,K\,\sinh^2y}\,\bigg(\frac{2\,[a_2 \cos(m\,y)+b_2 \sin(m\,y)]}{(K-1)\,\Psi(y)\,\textrm{sech}\,y}-1\bigg)\bigg].~~~~~~\label{18a}
\end{eqnarray}
\begin{figure*}[!htp]
   \centering
   \includegraphics[width =7cm]{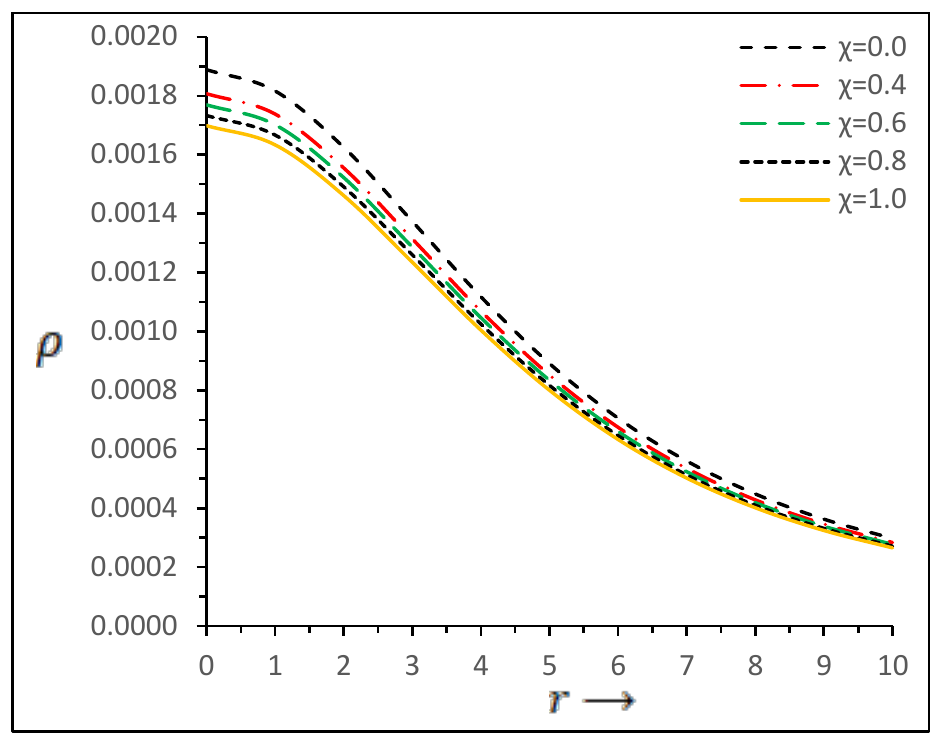} \includegraphics[width=7cm]{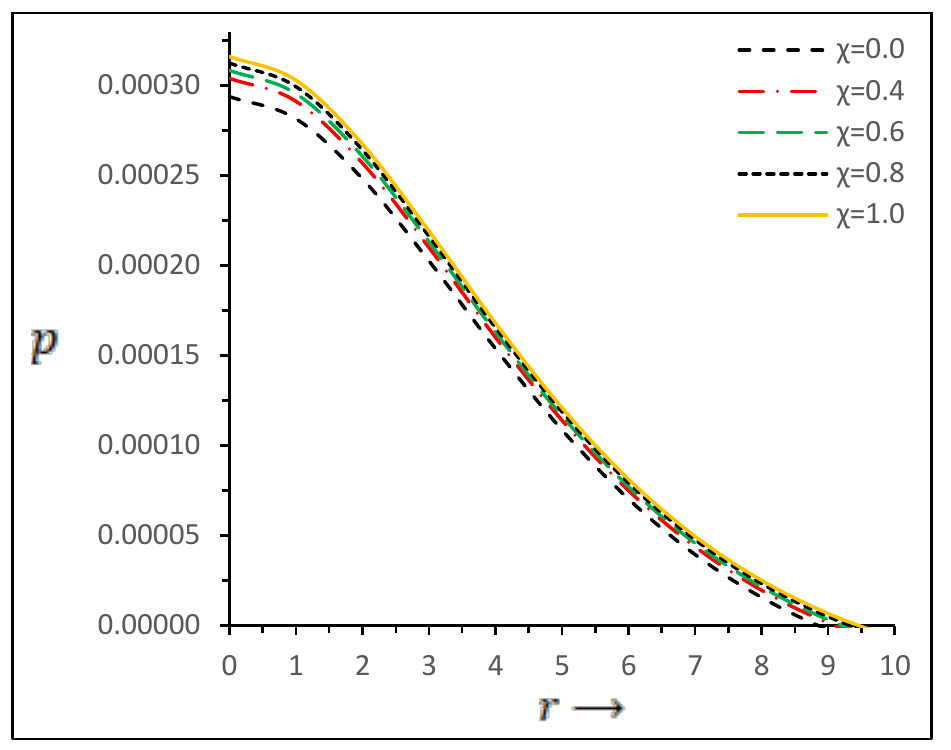}
   \caption{\label{fig5}The energy density (left panel) and pressure (right panel) diagram in model $f(R, T)$ = $R + 2 \chi T$ and in GR for compact stars with $K=3$, $M=1.3 M_{\odot}$ and $C =2.373\times 10^{-2}km^{-2}$. We find that for $\chi =0$ ( dashed black curve for GR case) the radius goes upto $R=8.849$ Km, whereas $\chi =0.4, 0.6, 0.8$ and $1$ the radius goes as high as $9.089$, $9.208$, $9.327$ and $9.411$, respectively.}
\end{figure*}
\begin{figure*}[!htp]
   \centering
   {\includegraphics[width=7cm]{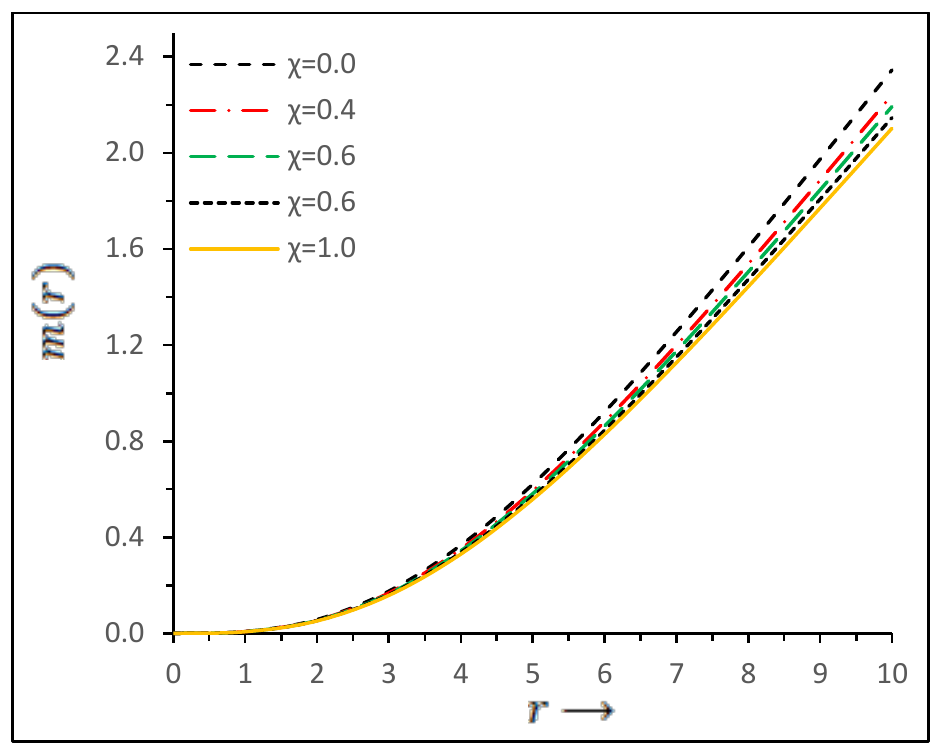} \includegraphics[width=7cm]{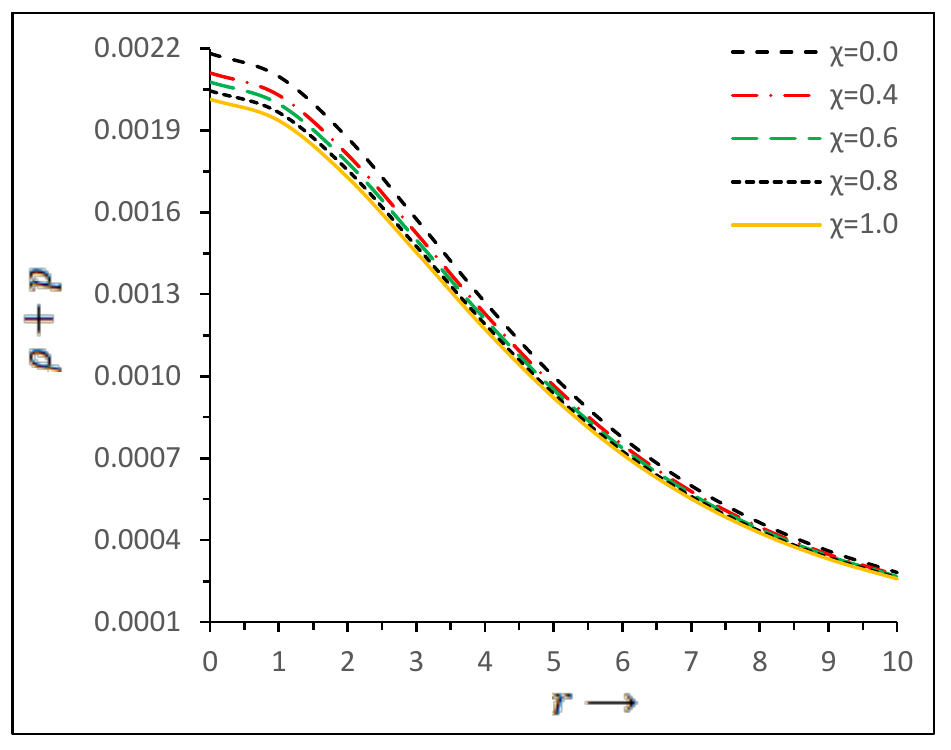}
\caption{\label{fig6} Gravitational mass $m(r)$ and Null energy condition (right diagram) versus radial coordinate $r$ have been plotted. The dashed black curves are the solutions of GR case, while the others for $f(R, T)$ model. From a rapid inspection of these plots, the differences between GR and $f(R, T)$ gravitational mass are clear and the tendency is that at larger radius GR takes more masses.}}
\end{figure*}
\begin{figure*}[!htp]
   \centering
   {\includegraphics[width=7cm]{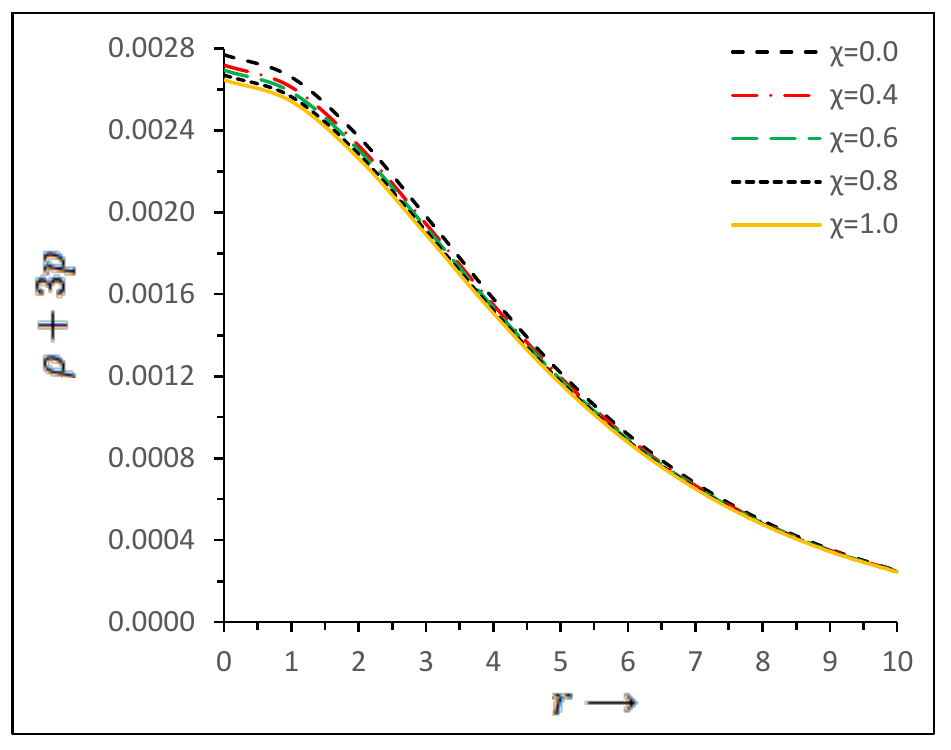} \includegraphics[width=7cm]{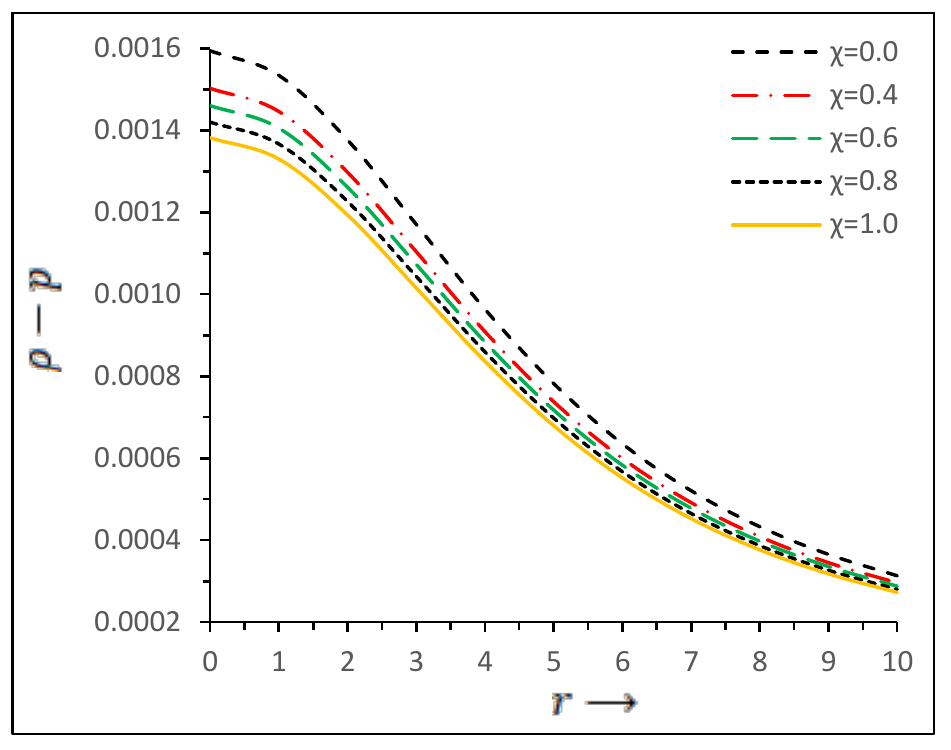}
\caption{\label{fig7}This diagram is for SEC and DEC against the radial coordinate $r$ for different chosen values of $\chi$. }}
\end{figure*}
The sound speed index is given by
\begin{eqnarray}
&&\hspace{-2cm}\left(\frac{dp}{d\rho}\right)_E=\frac{(K-1)\,\tanh y\, [a_2 \cos (m\,y) + b2 \sin (m\,y)]\, [2\,v_{21}(y) + \Omega_{E_{21}}(y)\big]}{[9 - K + (K-1) \cosh (2\,y)]\, \big[v_{21}(y) + \coth y\, \big(a_2 \cos (m y) + b_2 \sin (m y)\big)]^2},~~~~~~\\
&&\hspace{-2cm}\left(\frac{dp}{d\rho}\right)_f =\frac{\chi\,\big(2\,\coth y\, \textrm{csch} y - 3\, \Omega_ {f_ {21}} (y)\big) - (8 \pi+3 \chi )\, v_{25}(y) - \coth (y)\, v_{22}(y)}{(8 \pi+3 \chi )\,\big(2 \coth y\, \textrm{csch}  y - 3 \,\Omega_ {f_ {21}} (y) \big)- \chi\, v_{25}(y)- \coth (y) v_{23}(y)}.~~~~~
\end{eqnarray}

\begin{figure*}
   \centering
   {\includegraphics[width=7cm]{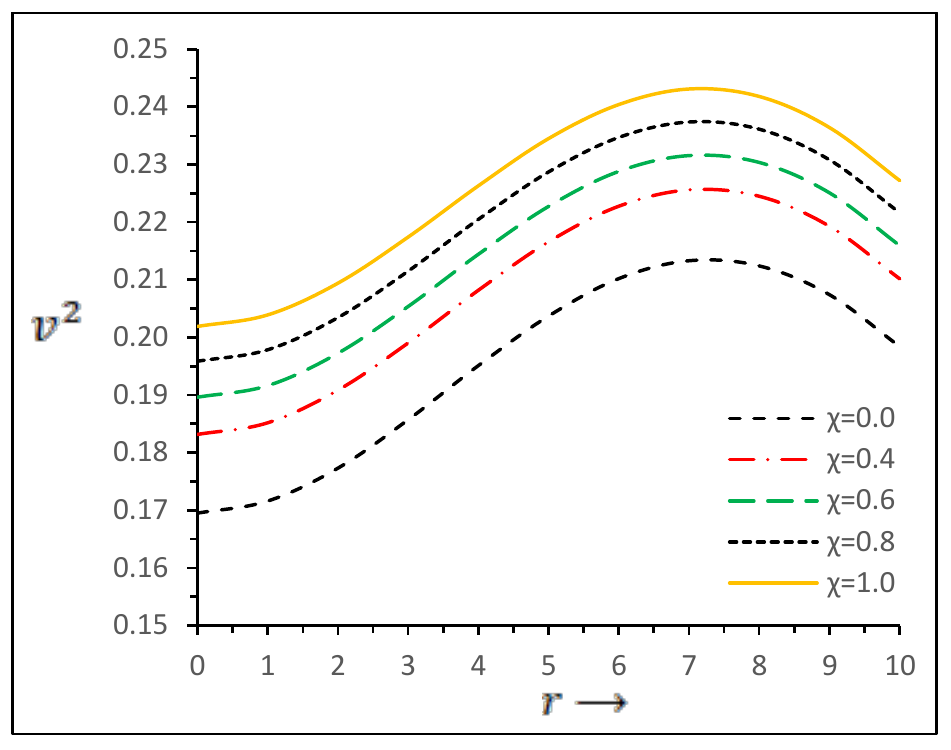}
\caption{\label{fig8} The sound speed obtained for parameters as on Fig. \ref{fig6}.}}
\end{figure*}
\begin{multicols}{2}
In this framework, let us now discuss the stellar structure with aim to study the physical validity and stability of the system under the $f(R,T)$ gravity.
The energy density and pressure versus radial distance from the center of the star have been plotted for each $\chi$ are depicted in Fig.~\ref{fig5}. As one can see that pressure and density for both Einstein and $f(R,T)$ model are maximum at the center and decreases monotonically towards the boundary. Fig.~\ref{fig5} confirms a well-behaved positive definite density. With the choice of the free parameters and depending  on matter content, one can increase and decrease the radius of the stellar structure.  For illustrating we assume that  $M=1.3 M_{\odot}$, the constant $C$= $ 2.373\times 10^{-2} km^{-2}$. According to the results we observe that for $\chi=0$ the radius goes upto $R=8.849$ Km, whereas $\chi =0.4, 0.6, 0.8$ and $1$ the radius goes as high as $9.089$, $9.208$, $9.327$ and $9.411$, respectively, where the pressure at the surface of the star is equal to zero $p (r = R) = 0$.
In Table \ref{table2}, we show the central density, central and surface pressure against the total radius for some different values of $\chi$. {Note that the energy density and pressure are in the same order of magnitude only near the centre of the star.}

In order to analyze the mass function (\ref{eq19}),
we integrate the system of equation considering many different values of $\chi$. We omit the mass expressions as they are extremely lengthy. Fig. \ref{fig6} shows the behavior of the total mass, normalized in solar masses versus the radius of the star. We find that variation in the parameters bring a significant changes to central density and pressure, without bringing any huge effect to mass-radius relation, see Table \ref{table2}.  From Fig. \ref{fig6} and Table \ref{table2}, we observe that the maximum gravitational mass $M=1.3 M_{\odot}$ attended from standard GR when $\chi$ = 0 with  radius $R=8.849$ km. {It is noteworthy from Table \ref{table2} that the modified effects of pressure on mass density and self gravity of the star are
extremely important when one seeks for high mass configurations of compact stars. At high mass, all the variations of $\chi$ become gradually ineffective and finally overlap with the normal GR solutions. This situation is also seen in \cite{Maurya:2018kxg}.}

Analyzing the Figs. \ref{fig6} and \ref{fig7}, our investigation shows that the energy conditions in both cases are well behaved inside the star. In Fig. \ref{fig8}, one can see the sound speed in both cases are satisfied. Surprisingly the maximum velocity, as seen from the models, is lowest for ordinary GR (when $\chi$ =0).The value of parameters are enlisted in Figs. \ref{fig5}.
\end{multicols}
{\centering{\subsubsection{\textbf{Case IIIb: $2-K= m^2$,~~\textrm{and}~~$1<K<2$}}}}
Let us consider the model for $1<K<2$, and compare the obtained results with $K >2$. Under these assumptions, Eqs. (\ref{12b}) and (\ref{hyper}) lead to the following set of equations
\begin{eqnarray}
&&\rho_E =\frac{C\,[ 3-K+(K-1)\,\cosh^2y ]}{8\,\pi\,K\,(K-1)\,\sinh^4y} ,~~\label{dE2b}\\
&&p_E =\frac{C}{8\pi K\sinh^2y}\bigg[\frac{2[c_2\cosh(m\,y) + d_2\sinh(m\,y)]}{(K-1)\,\textrm{sech}\,y\,\Psi(y)}-1\bigg].\label{pE2b}\\
 &&\rho_f = \frac{C}{(8\pi^2+6\pi\chi+\chi^2)}\bigg[\frac{(8\pi+3\chi)[ 3-K+(K-1)\cosh^2y ]}{8\,K\,(K-1)\,\sinh^4y}+\frac{\chi}{8\,K\,\sinh^2y}\,\bigg(\frac{2\,[c_2\,\cosh(m\,y) + d_2\, \sinh(m\,y)]}{(K-1)\,\textrm{sech}\,y\,\Psi(y)}-1\bigg)\bigg],~~~~~ \label{17b}\\
 &&p_f= \frac{C}{(8\pi^2+6\pi\chi+\chi^2)}\,\bigg[\frac{\chi\,[ 3-K+(K-1)\,\cosh^2y ]}{8\,K\,(K-1)\,\sinh^4y} +\frac{(8\pi+3\chi)}{8\,K\,\sinh^2y}\,\bigg(\frac{2\,[c_2\,\cosh(m\,y) + d_2\,\sinh(m\,y)]}{(K-1)\,\textrm{sech}\,y\,\Psi(y)}-1\bigg)\bigg].~~~ \label{18b}
\end{eqnarray}
\begin{figure*}[!htp]
   \centering
   {\includegraphics[width = 7cm]{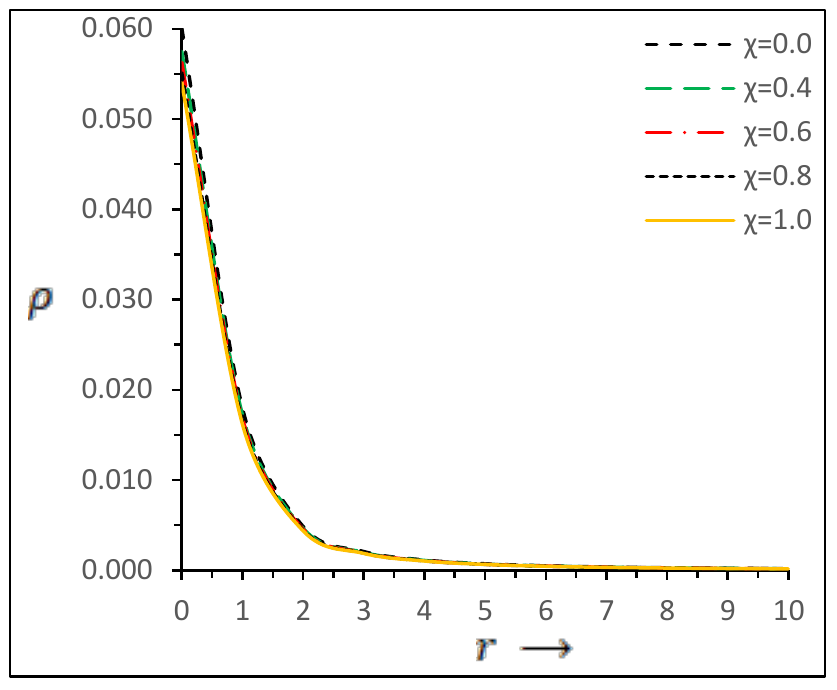}~~~ \includegraphics[width=7cm]{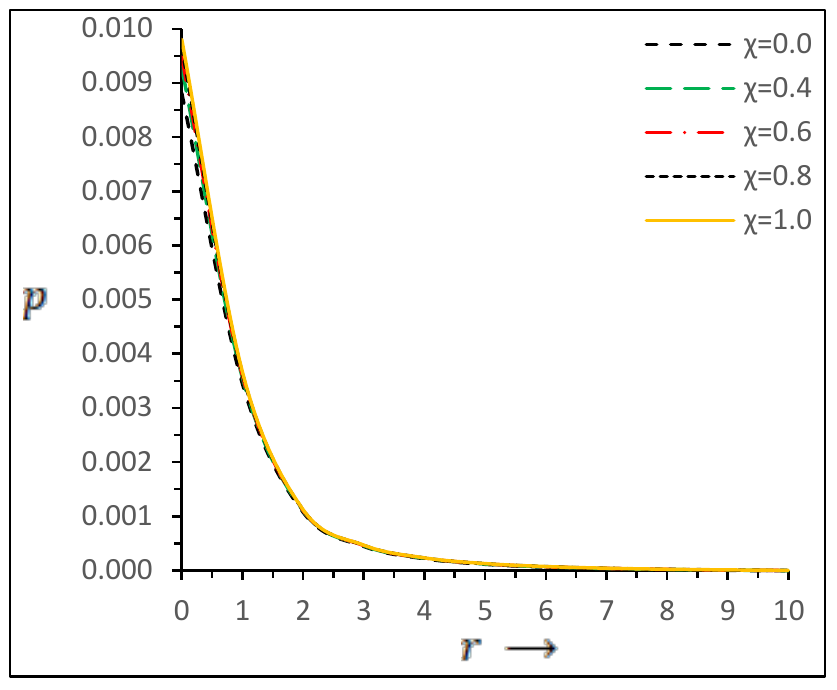}
   \caption{\label{fig9} The energy density and pressure diagram in model $f(R, T)$ gravity and in GR for compact stars with $K= 1.78$, $M=1.3 M_{\odot}$ and $C =1.1493$ $km^{-2}$. We find that for $\chi = 0$ ( dashed black curve for GR case) the radius goes upto $R=8.849$ Km, whereas $\chi =0.4, 0.6, 0.8$ and $1$ the radius goes as high as $9.242$,$9.437$, $9.632$ and $9.827$, respectively. }}
\end{figure*}
\begin{figure*}
   \centering
   \includegraphics[width=7.4cm]{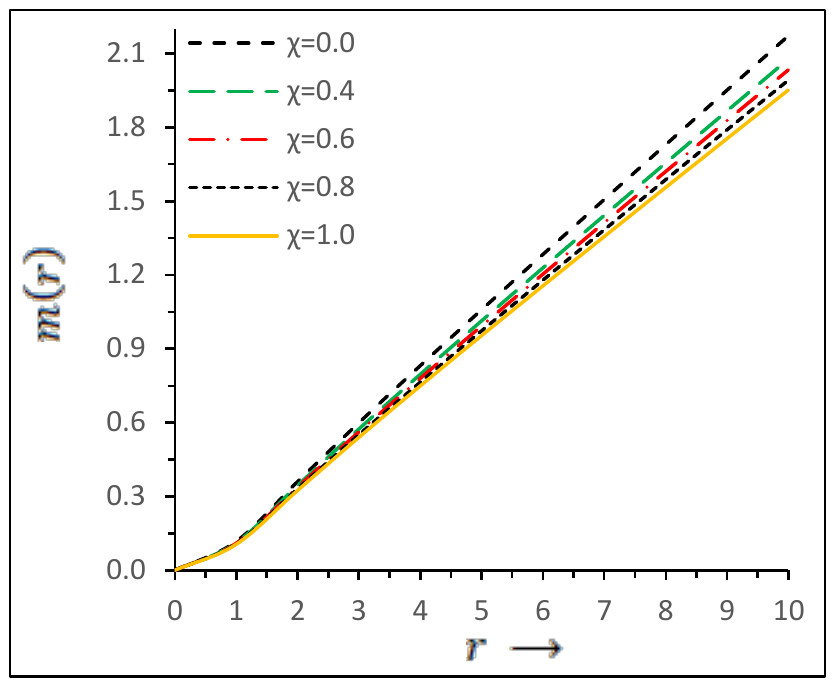}~~~\includegraphics[width=7.4cm]{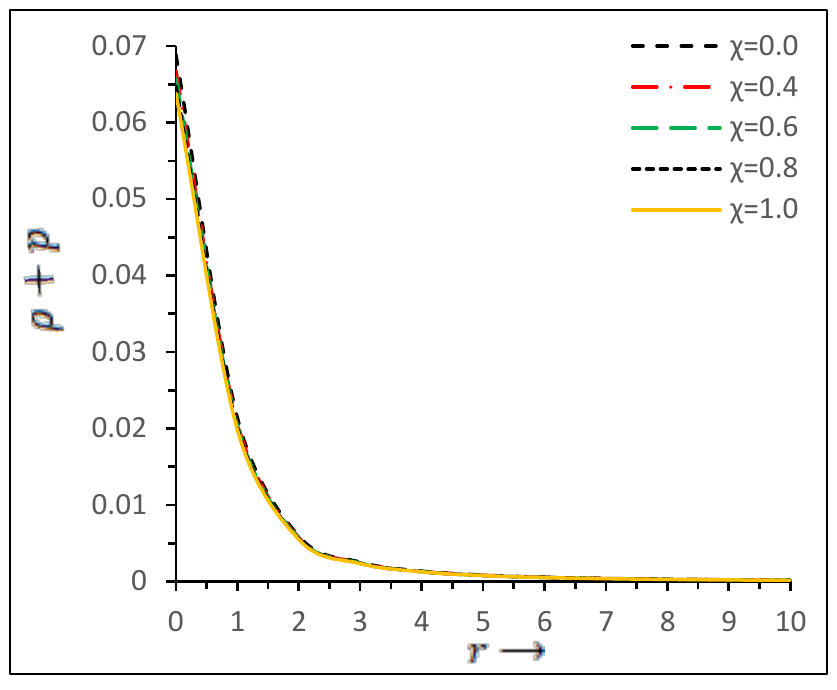}
\caption{\label{fig10} Variation of mass (top) and null energy condition  (bottom) versus radial coordinate $r$ have been plotted. We have used the same data set as of Fig. \ref{fig9}.}
\end{figure*}
\begin{figure*}
   \centering
   \includegraphics[width=7.4cm]{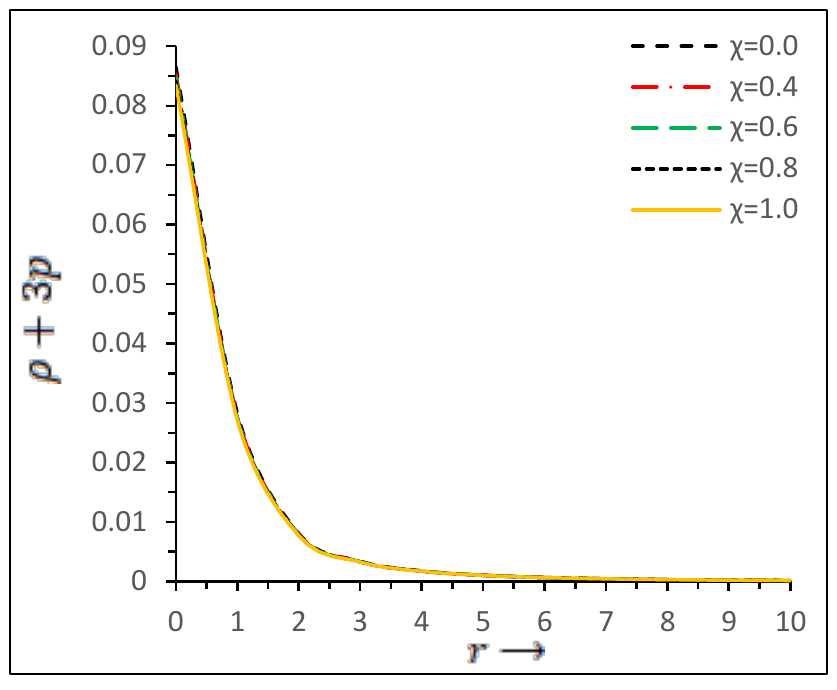}~~~ \includegraphics[width=7.4cm]{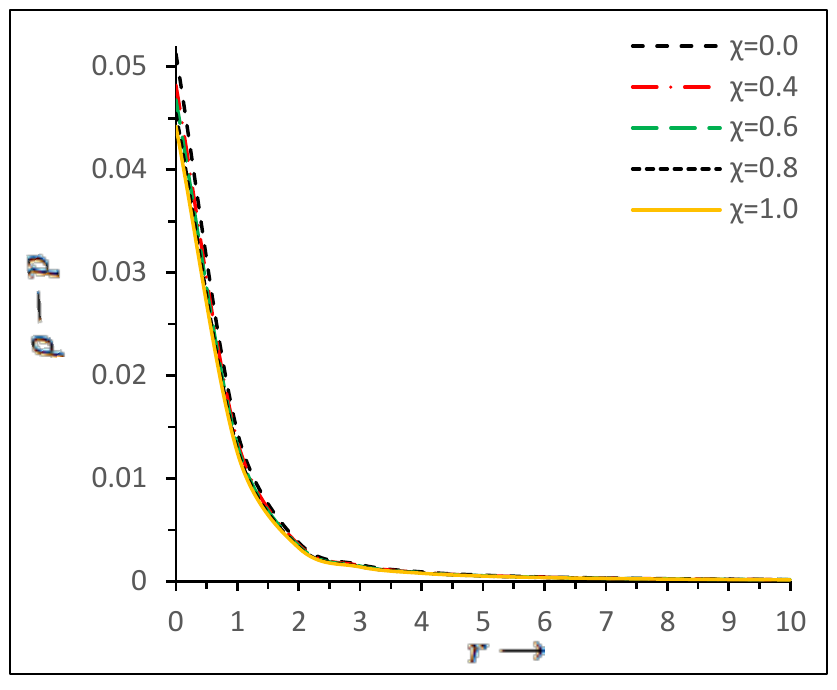}
\caption{\label{fig11} This diagram is for SEC and DEC against radial coordinate $r$  for different chosen values of $\chi$.}
\end{figure*}
\begin{figure*}
   \centering
   {\includegraphics[width=7.5cm]{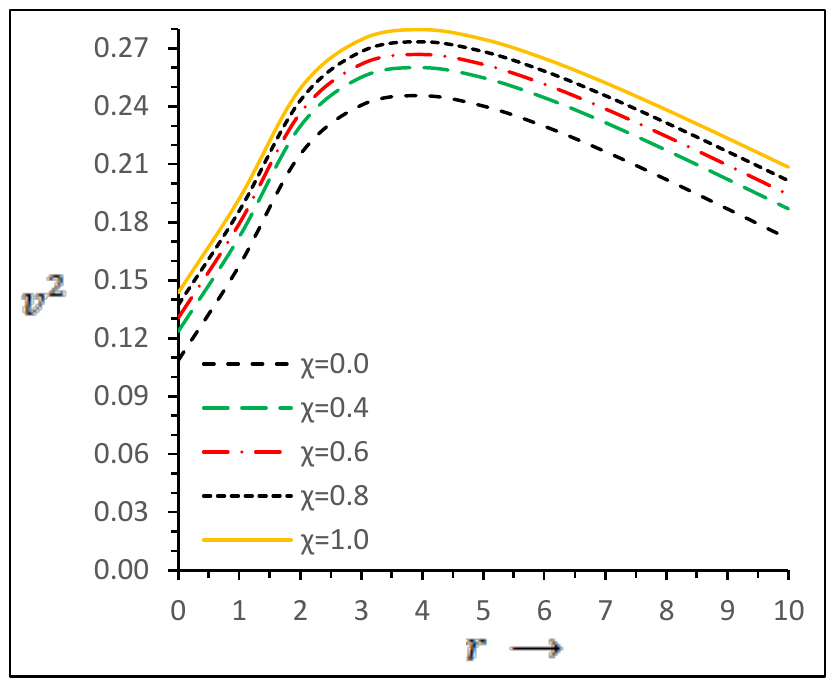}
\caption{\label{fig12} The sound speed obtained for parameters as on Fig. \ref{fig9}.}}
\end{figure*}

and the sound speed index inside the fluid is given by
\begin{eqnarray}
&&\left(\frac{dp}{d\rho}\right)_E =\frac{2\,(1 - K)\, \big[\Omega_ {E_ {22}} (y) + c_2^2 m \cosh(m\,y)\,\textrm{csch}^2 y\,\sinh(m\,y) - d_2\, \Omega_ {E_ {23}} (y)\big]}{\textrm{csch}^4y\,[9 - K + (K-1) \cosh(2\,y)\,\coth y\,]\,[\Omega_ 5 (y)]^2 },~~~~~\\
&&\left(\frac{dp}{d\rho}\right)_f =\frac{ \chi\,v_{31}(y) + (8\pi+3\chi)\, [\Omega_{f_{31}}(y)+ \Omega_{f_ {32}}(y)]-\coth y\, v_{32}(y)}{(8\pi+3\chi)\,v_{31}(y) + \chi\, [\Omega_{f_{31}}(y) + \Omega_{f_ {32}}(y)] -\coth y\, v_{33}(y)}.~~~~
\end{eqnarray}
\begin{multicols}{2}

For the study of compact objects,  Fig. \ref{fig9} depicts the energy density and isotropic pressure, respectively. We find that density and pressures are monotonically decreasing functions of the radial variable $r$ which is expected from Buchdahal model of GR \cite{Maurya:2018kxg}. Comparing $f(R,T)$ gravity with the previous model ( $K>2$), one can see that both  density and pressures decreases more rapidly towards the boundary. From a mathematical point of view, the difference between energy density at centre and surface are of order $\backsim 10^{2}$ magnitude only near the centre of the star (see Table \ref{table3}). Comparing both the Table \ref{table2} and \ref{table3}, our conclusion is that for a given radius with $1<K<2$ the star is most compact, as the central density is of order $\sim$ $10^{15} gm/cm^3$. The results reported for a particular mass (in normalized form)
$M=1.3 M_{\odot}$ with the constant $C =1.1493$ $km^{-2}$. For larger value of $\chi =1$ the convergence is slower and the radius goes upto $9.827$, as of Fig. \ref{fig9}. Comparing this results with our previous model, one can observe that gravitational mass increases in a rate with much faster than $K>2$ model. It is noticeable that the mass-radius relation differs significantly from the corresponding case in GR for reasonably high value of $\chi$. {We furthermore recall the standard results of GR from our previous paper \cite{Maurya:2018kxg}, and its solutions for $K=1.78$,  which can be compared to the analogous solutions coming from the modified gravity.  It is possible to demonstrate that density radial profiles coming from $f(R,T)$ gravity analytic models and close to those coming from GR are compatible.}
The results reported in Fig. \ref{fig10} and Fig. \ref{fig11} are obtained for different values of $\chi$, for the energy conditions.
In figures it is observed that all energy conditions are obeyed. The variation of square of sound speed is displayed in Fig. \ref{fig12}. It is also evident that the square of sound speed is less than unity throughout the stellar interior.\\
\end{multicols}
{\centering{\subsubsection{\textbf{Case IIIc: $2-K=0$}}}}

We next analyze the result for $K=2$. The density and pressure for the Einstein as well as $f(R, T)$ gravity are given by
\begin{eqnarray}
&&\rho_E=\frac{C\,[ 3-K+(K-1)\,\cosh^2y ]}{8\,\pi\,K\,(K-1)\,\sinh^4y} ,~~\label{eq46} 
\\
&& p_E =\frac{C}{8\,\pi\,}\bigg[\frac{2\,\cosh y\,(f_2\,y+e_2)-(K-1)\,\Psi(y)}{K\,(K-1)\,\sinh^2y\,\Psi(y)}\bigg],\label{eq47}\\
&& \rho_f = \frac{C}{(8\pi^2+6\pi\chi+\chi^2)}\bigg[\frac{(8\pi+3\chi)[3-K+(K-1)\cosh^2y ]}{8\,K\,(K-1)\,\sinh^4y}+\frac{\chi}{8\,K}\,\bigg(\frac{2\,\cosh y\,(f_2\,y+e_2)-(K-1)\,\Psi(y)}{(K-1)\,\sinh^2y\,\Psi(y)}\bigg)\bigg],\label{eq48}\\
&& p_f = \frac{C}{(8\pi^2+6\pi\chi+\chi^2)}\bigg[\frac{\chi\,[3-K+(K-1)\,\cosh^2y ]}{8\,K\,(K-1)\,\sinh^4y}+\frac{(8\pi+3\chi)}{8\,K}\,\bigg(\frac{2\,\cosh y\,(f_2\,y+e_2)-(K-1)\,\Psi(y)}{(K-1)\,\sinh^2y\,\Psi(y)}\bigg)\bigg].\label{eq49}~~~~~~~
\end{eqnarray}
\begin{figure*}[!htp]
   \centering
   {\includegraphics[width = 7cm]{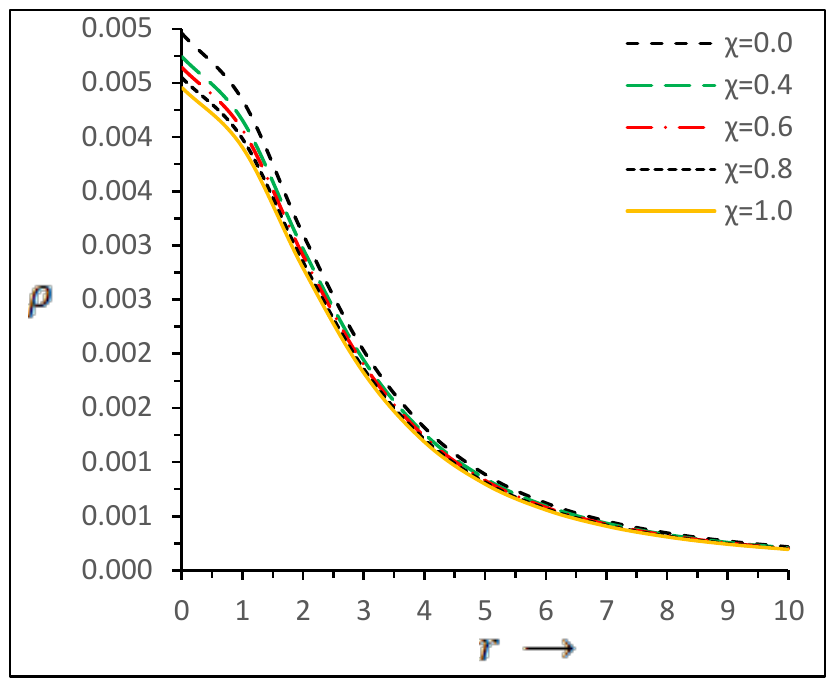}~~~\includegraphics[width=7cm]{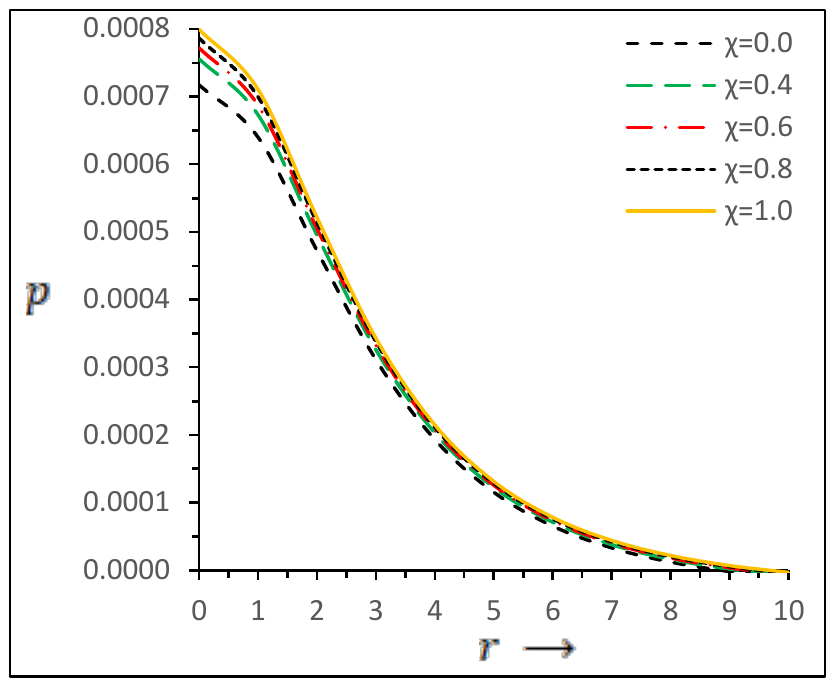}
   \caption{\label{fig13}The energy density and pressure versus radial coordinate $r$  have been plotted for different values of $\chi$ and fixed positive value of $K=2$. For plotting we consider $\chi =0$ ( dashed black curve for GR case) reveals that for a mass of $M=1.3 M_{\odot}$, the radius goes upto $R=8.849$ Km, whereas $\chi =0.4, 0.6, 0.8$ and $1$ the radius goes as high as $9.171$,$9.331$, $9.491$ and $9.651$, respectively. In all cases, it is considered the constant $C =0.831$ $km^{-2}$. }}
\end{figure*}

\begin{figure*}
   \centering
   {\includegraphics[width=7cm]{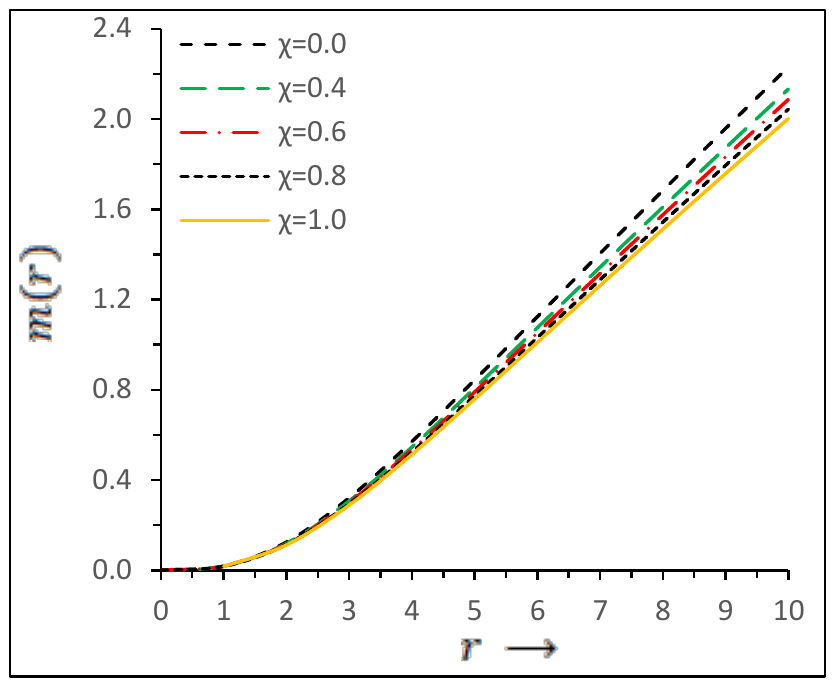}~~~\includegraphics[width=7cm]{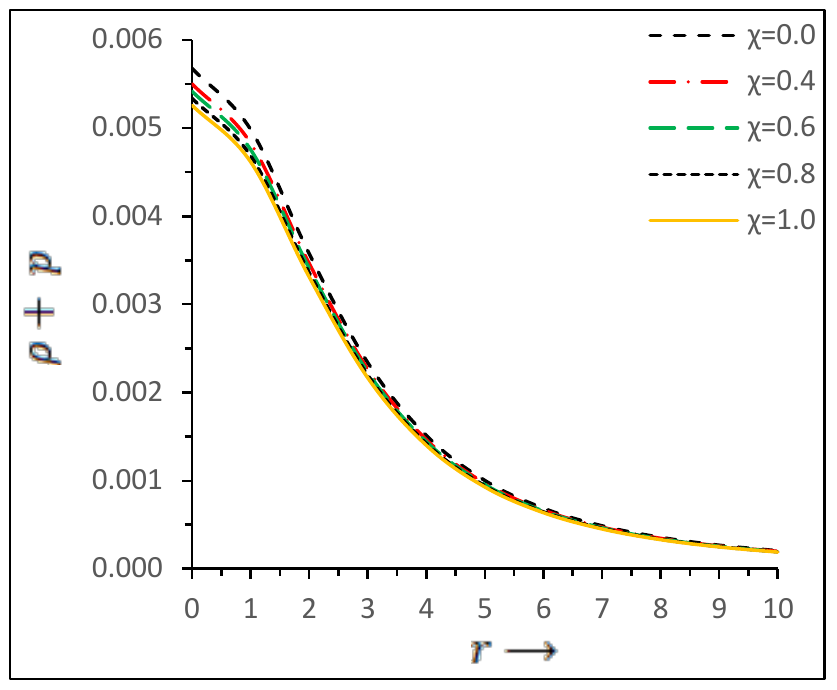}
\caption{\label{fig14}Mass-radius relations in the GR and in the $f(R,T)$ gravity are shown. The NEC (right diagram) is determined by the condition $\rho+p >0$. We consider the same data set as of Fig. \ref{fig13}.}}
\end{figure*}
\begin{figure*}[!htp]
   \centering
   {\includegraphics[width=7cm]{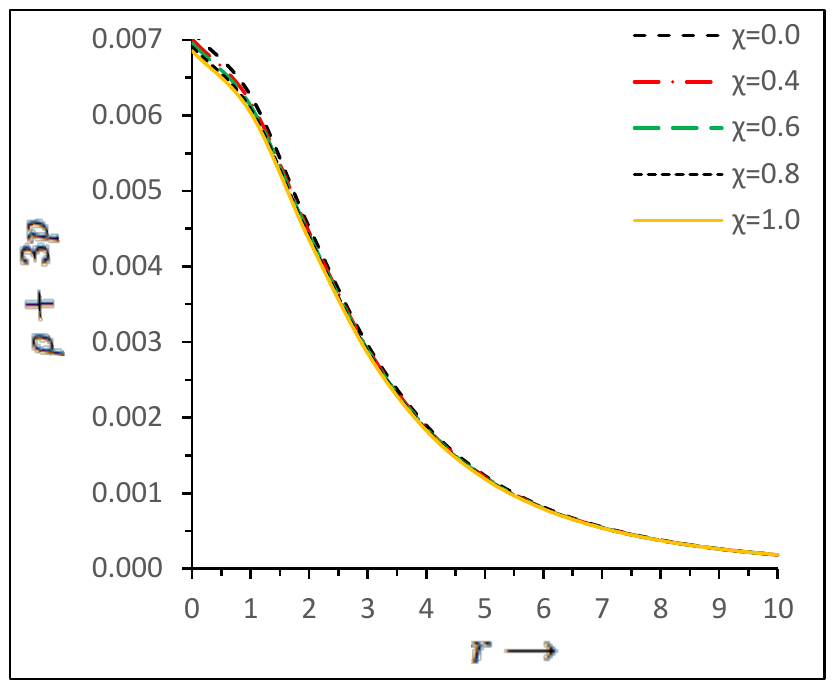}~~~ \includegraphics[width=7cm]{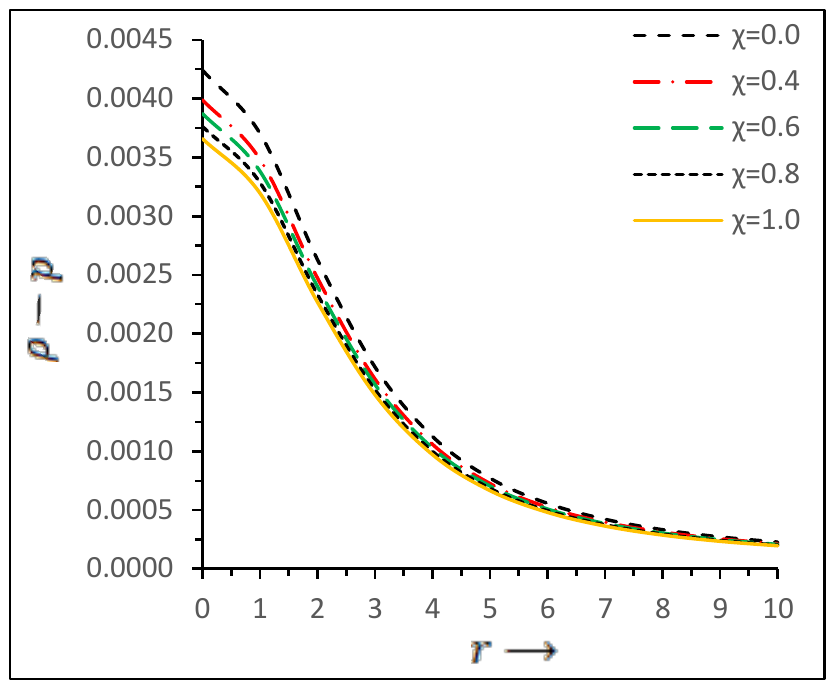}
\caption{\label{fig15}Variation of the  SEC and DEC versus radial coordinate $r$ have been plotted for  $\chi =0.4, 0.6, 0.8$ and 1. }}
\end{figure*}
The square of the acoustic speed $dp/ d\rho$ becomes
\begin{eqnarray}
&&\hspace{-2cm}\left(\frac{dp}{d\rho}\right)_E =\frac{2\,[2\, (2-K)\,f_2\,(e_2 + f_2 y)\,\coth^2 y  + \Omega_ {9}(y)]}{[9 - K + (K-1) \cosh (2\,y)]\,\coth y\, \textrm{csch}^4 y\, v_{41}(y)},~~ \label{eq50}\\
&&\hspace{-2cm}\left(\frac{dp}{d\rho}\right)_f=\frac{2 \chi\,[(K-1) \coth y -2\Omega_ {10} (y)] + (8\pi+3\chi)[\,v_{42}(y)+2\, \coth y \,\Omega_ {f_ {43}} (y)]} {2\,(8\pi+3\chi)\,[(K-1)\, \coth y\, -2 \Omega_ {10} (y)] + \chi\,v_{42}(y) +2\chi \coth y\, \Omega_ {f_ {43}} (y)]}. \label{eq51}
\end{eqnarray}
\begin{figure*}[!htp]
   \centering
   {\includegraphics[width=7cm]{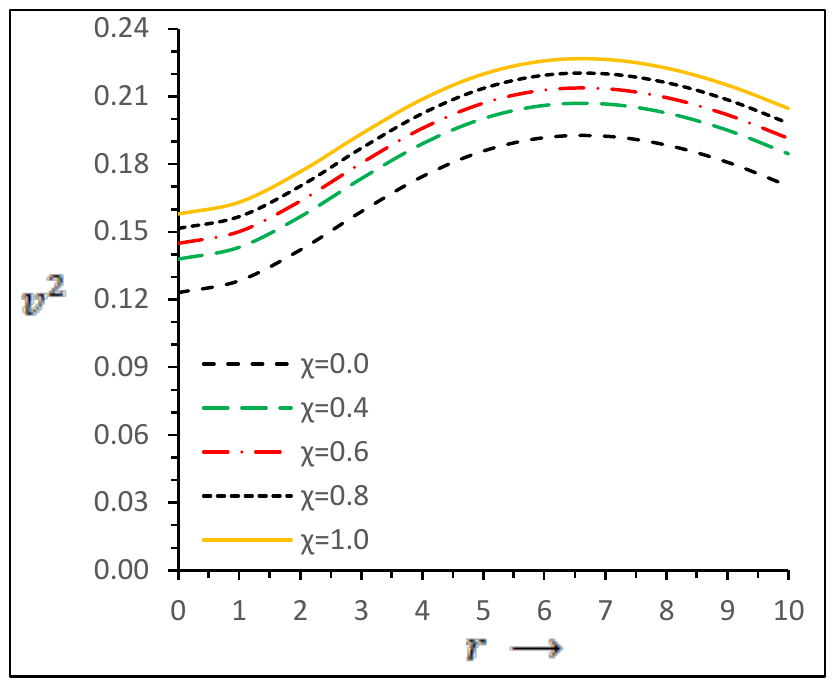}
\caption{\label{fig16} Variation of square of the sound velocity $v^2$
 versus radial coordinate $r$  for $\chi =0$ (dashed black curve) $\chi =0.4$ (dash-dotted red curve), $\chi =0.6$ (long dashed green curve), $\chi =0.8$ (small dashed black curve) and $\chi =1$ (yellow curve). }}
\end{figure*}
\begin{multicols}{2}
 Finally, we move on to describe the results obtained from our calculations, which are illustrated
in Fig. \ref{fig13}, for $K$ = 2 and for different values of $\chi$. We notice that the solution gives positive definite density and pressure which is also monotonic decreasing towards the boundary.  In Fig. \ref{fig14}, we have depicted the variations of mass with respect to the distance from the center of a  star. Notice that the differences between GR and $f(R,T)$ gravitational mass is clear from the figures and the tendency is that at progressively decreasing values of $\chi$,  $f(R,T)$ model acquire more mass but always less than GR. Considering  Fig. \ref{fig13} and \ref{fig14}, and the above Eqs. (\ref{eq46}-\ref{eq49}), our results are given in Table \ref{table4}.
 According to Fig. \ref{fig15}, we observe that all energy conditions are satisfied.
 So, our assumption for star model introduced in this paper is suitable. Furthermore, the speed of sound is defined through the Eqs. (\ref{eq50}-\ref{eq51}),
 and presented at Fig. \ref{fig16}. We see from this figure that the speed of sound is less than unity throughout the star.
In conclusion we explore all the mathematical and graphic analysis of our study. 
In such situations, a couple of comments are necessary. Regarding the numerical values tabulated in Table (\ref{table1}-\ref{table4}), however,
the results indicate that GR dominates all the results obtained in the $f(R,T)$ scenario.
As consequence, in the framework of GR, the objects present a denser core than in the $f(R,T)$ theory,
and therefore greater value of gravitational redshift $Z_{s}$ do exists. In summary, from Tables (\ref{table1}-\ref{table4}), the tendency of the main physical parameters that describe the micro-physical behavior (central energy-density, surface energy-density, central pressure) of the system tend to decreasing in magnitude when the parameter $\chi$ increase its value.

\begin{table*}[!htp]
\centering
\caption{\label{table1}Comparative study of physical values of the compact star in $f(R, T)$ gravity and GR for $C=1.8047\times 10^{-3} Km^{-2}$, mass $=1.3 M_{\odot}$ and $K= -0.4$ for different $\chi$.}
\begin{tabular}{ccccccc}\hline
 {$\chi$}& Radius  & Central pressure &
Central density & Surface density & $M-R$ ratio  & Surface red\\
& $R (km)$ & $p_c^{eff}$ ($dyne/cm^2$) & $\rho_c^{eff} (gm/cm^3$) & $\rho_s^{eff} ( gm/cm^3$) &$\frac{2M}{R}$&-shift ($Z_s$) \\ \hline
0.0 &  8.849  & 6.86953$\times10^{33}$ & 4.04882$\times10^{13}$ & 3.25466$\times10^{13}$ & 0.433382 & 0.32846\\
0.4 & 8.997 & 6.60441$\times10^{33}$ & 3.87546$\times10^{13}$ & 3.08524$\times10^{13}$ & 0.42626 & 0.343669\\
0.6 & 9.069 & 6.48504$\times10^{33}$ & 3.79428$\times10^{13}$ & 3.00643$\times10^{13}$ & 0.42286 & 0.35131\\
0.8 & 9.141 & 6.37628$\times10^{33}$ & 3.71647$\times10^{13}$& 2.93103$\times10^{13}$ & 0.41954 & 0.35910\\
1.0 & 9.211 & 6.27274$\times10^{33}$ & 3.64184$\times10^{13}$ & 2.85910$\times10^{13}$ & 0.41635 & 0.36684\\ \hline
\end{tabular}

\end{table*}
\begin{table*}[!htp]
\centering
\caption{\label{table2}Comparative study of physical values of the compact star in $f(R,T)$ gravity and GR for $C=2.373\times 10^{-2} Km^{-2}$, mass $=1.3 M_{\odot}$ and $K=3$ for different $\chi$.}
\begin{tabular}{ccccccc}\hline
 {$\chi$}& Radius  & Central pressure &
Central density & Surface density & $M-R$ Ratio & Surface \\
& $R (km)$ & $p_c^{eff}$ ($dyne/cm^2$) & $\rho_c^{eff}$ ($gm/cm^3$) & $\rho_s^{eff}$ ($gm/cm^3$) &$\frac{2M}{R}$& redshift ($Z_s$) \\ \hline
0.0 &  8.849 & 1.42038$\times10^{34}$ & 1.01406$\times10^{14}$ & 2.01018$\times10^{13}$  & 0.433382 & 0.32852\\
0.4 & 9.089 & 1.46946$\times10^{34}$ & 0.9703249$\times10^{14}$ & 1.82605$\times10^{13}$ & 0.42194 & 0.33805\\
0.6 & 9.208 & 1.49081$\times10^{34}$ & 0.949973$\times10^{14}$ & 1.74261$\times10^{13}$ & 0.41648 & 0.34271\\
0.8 & 9.327 & 1.51038$\times10^{34}$ & 0.930537$\times10^{14}$ & 1.66410$\times10^{13}$ & 0.41116 & 0.34731\\
1.0 & 9.411 & 1.52231$\times10^{34}$ & 0.911931$\times10^{14}$ & 1.601346$\times10^{13}$ & 0.4075 & 0.35053\\\hline
\end{tabular}
\end{table*}

\begin{table*}[!htp]
\centering
\caption{\label{table3}Comparative study of physical values of the compact star in $f(R,T)$ gravity and GR for $C=1.1493 Km^{-2}$, mass $=1.3 M_{\odot}$ and $K=1.78 Km$ for different $\chi$.}
\begin{tabular}{ccccccc}\hline
 {$\chi$}& Radius  & Central pressure &
Central density & Surface density & $M-R$ Ratio & Surface red-\\
& $R (km)$ & $p_c^{eff}$ ($dyne/cm^2$) & $\rho_c^{eff}$ ($ gm/cm^3$) & $\rho_s^{eff}$ ( $ gm/cm^3$) &$\frac{2M}{R}$& shift ($Z_s$)\\ \hline
0.0 &  8.849 & 5.26129$\times10^{35}$  & $3.22822\times10^{15}$  & 1.20854$\times10^{13}$ & 0.433382 & 0.32848 \\
0.4 & 9.242 & 4.49111$\times10^{35}$  & $3.08869\times10^{15}$ & 1.05655$\times10^{13}$ & 0.41496 & 0.32895\\
0.6 & 9.437 & 4.58570$\times10^{35}$  & $3.02382\times10^{15}$ &  0.990379$\times10^{13}$ & 0.40638 & 0.32916 \\
0.8 & 9.632 & 4.66868 $\times10^{35}$ & $2.96189\times10^{15}$ &  0.929627$\times10^{13}$ & 0.39816 & 0.32936 \\
1.0 & 9.827 & 4.74127$\times10^{35}$  & $2.90269\times10^{15}$ &  0.873750$\times10^{13}$ & 0.35926 & 0.32954\\\hline
\end{tabular}
\end{table*}

\begin{table*}[!htp]
\centering
\caption{\label{table4}Comparative study of physical values of the compact star in $f(R, T)$ gravity and GR for $C=0.831\times 10^{-1}$ Km,  mass $=1.3 M_{\odot}$ and $K=2$ for different $\chi$.}
\begin{tabular}{ccccccc}\hline
 {$\chi$}& Radius  & Central pressure &
Central density & Surface density & $M-R$ ratio & Surface red-\\
& $R (km)$ & $p_c^{eff}$ ($dyne/cm^2$) & $\rho_c^{eff}$ ($gm/cm^3$) & $\rho_s^{eff}$ ( $ gm/cm^3$) &$\frac{2M}{R}$& shift ($Z_s$) \\ \hline
0.0 &  8.849  & 3.46718$\times10^{34}$ & 2.66334$\times10^{14}$ &  1.49764$\times10^{13}$ & 0.433382 &  0.32850\\
0.4 & 9.171 & 3.65329$\times10^{34}$ & 2.54813$\times10^{14}$ &  1.32607$\times10^{13}$ & 0.41816 & 0.33323\\
0.6 & 9.331 & 3.73173$\times10^{34}$ & 2.49457$\times10^{14}$ &  1.25027$\times10^{13}$ & 0.411 & 0.33545\\
0.8 & 9.491 & 3.80177$\times10^{34}$ & 2.44345$\times10^{14}$ & 1.18012$\times10^{13}$ & 0.40406 & 0.33759\\
1.0 & 9.651 & 3.8642$\times10^{34}$ & 2.39459$\times10^{14}$ & 1.11509$\times10^{13}$ & 0.39736 & 0.33965\\\hline
\end{tabular}
\end{table*}

\section{Final Remarks}\label{sec5}
{To conclude our study, we highlight the most important results from the obtained solutions. First of all, it is worth mentioning that Buchdahl model \cite{Buchdahl:1959rhi} describing a static spherically symmetric spacetime driven by an isotropic matter distribution is extended into the framework of modified gravity theories, specifically $f(R,T)$ gravity. To accomplish it, we have considered to main ingredients. The first one is $f(R,T)=R+2\chi T$ \i.e. a linear function in both $R$ and $T$, where $\chi$ is a dimensionless running parameter. The second one refers to the form of  energy-momentum tensor $T_{\mu\nu}$. In this respect we have taken a perfect fluid matter distribution. The foundations of the mentioned elections lies on the following statements: i) the linear $f(R,T)$ function allows a more tractable mathematical treatment, additionally as  pointed out earlier that extra piece $2\chi T$ can be consolidated as a running cosmological constant. ii) Einstein gravity theory and $f(R,T)$ gravity share the same isotropic condition (\ref{iso}). In other words, any perfect fluid solution to Einstein field equations is also a solution in the arena of $f(R,T)$ gravity. Of course, it is only from the mathematical point of view, since the material content is quite different. Precisely, this feature is the key starting point to address cosmological issues from the perspective of compact structure within the astrophysical framework.
All things mentioned above allow us to examine the viability of Buchdahl models in the framework of $f(R,T)$ gravity and compare the results with pure GR solutions.} In this respect, in our previous article, we examine the Buchdahl model for anisotropic fluid sphere \cite{Maurya:2018kxg} and showed that one can obtain an analytic solution to the Einstein equations for positive and negative values of Buchdahl parameter $K$. 

So, depending on the parameters of the model $K$ and $\chi$, we analyse the configuration from $\chi = 0$ to $\chi$  reasonably high. As applied to compact non-rotating stars, the field equations are solved by applying  \textit{Gupta-Jasim} two step method (see review for details discussion \cite{Maurya:2018kxg,Kumar:2018hgm}). Here we report some progress in this direction:
we derive modified field equations and address issues regarding the choice of $\chi$. This has been done for two classes of models: the negative $K$ and positive $K$. The obtained results are summarized as follows:

\begin{itemize}
    \item In our analytical approach for $K <0$, we found that our stellar model is free from any geometrical singularities. In figs. \ref{fig1}-\ref{fig4}, we have shown all the physical criteria for stellar structure as described in the introduction. The results of calculations are given in Table \ref{table1}. We found that central energy density and surface density is much consistent for both GR and modified gravity with same magnitude. Also Table \ref{table1} exhibits the central pressure, the surface redshift and the mass-radius ratio for our predicted values. We find that the radius of the star increase as the coupling parameter $\chi$ increases, but the maximal mass limit
    exists for $\chi=0$. In conclusion, taking $\chi=0$ i.e. for GR the object becomes more compact, dense and massive.

    \item In next, we have analysed the model for $K >2$. Depending on the physical parameters, we have plotted Figs. \ref{fig5}-\ref{fig8} together with the solution of the classical GR ($\chi=0$). Independently from Table \ref{table2} one can see the decreasing value of the central and surface density, central pressure, surface redshift and the value of $2M/R$, with the increasing value of $\chi$. We note here that variations in the parameters does not effect huge to mass-radius relation, however, it increases monotonically towards the boundary, as of Fig. \ref{fig6}.

    \item For further precision we analyzed the model for $1< K <2$.
    The mass-radius relation and the variation of density and pressure with radius of the star for this $f(R,T)$ model are depicted in Figs. \ref{fig9}-\ref{fig12}.
    It is worth mentioning that the difference between central density and surface density is very high (in Table \ref{table3}), which is also higher than any other comparative model. In the case $\chi=0$, we have the central pressure is about 5.7 $\times 10^{35}$ orders of magnitude which is larger than the $\chi=1$ for 4.7$\times 10^{35}$. It is clear that the masses and radii of the stars change with the increasing values of $\chi$.

    \item Finally, in Figs. \ref{fig13}-\ref{fig16} and Table \ref{table4} we show the effects of $f(R, T)$ theory in compact star properties obtained for $ K =2$. As evident form
     Fig. \ref{fig14} that, for relatively small value of $\chi$ the model acquire more masses but these value is less than GR. Analyzing the maximum mass and its respective radius found in each curve we determine that these values could change from 2\% to 10\%. Also, for small value of $\chi$, the velocity of sound is less than unity.
\end{itemize}

For the sake of comparison with the results in the literature, we arrive at conclusion that among the four models, it is notable that, for $1<K<2$, we obtain the maximum density of order $10^{15} gm/cm^3$ and pressure $10^{35}dyne/cm^2$ of stars.
Our considerations show that for describing $\chi$, the total gravitational mass increases and reach at maximum when $\chi=0$ i.e. GR case. The radii of spheres are around around nine Km for realistic values of $\chi$. It appears that self gravity has more pronounced effect on the gravitational mass because high mass configurations are obtained only when $\chi= 0$. {Furthermore, we see that the general trend predicts a larger radius of a compact star corresponding to the higher values of $\chi$.  Increasing the values of $\chi >0$ may increase the radius of a star, but the mass in $f(R,T)$-gravity is lower than the mass in GR.
On the other hand, we conclude that the Buchdahal model gives a well behaved solution for all values of $K<0~~ \textrm{and}~~ K>1$ in GR as well as $f(R,T)$ gravity theory.  But unfortunately, both model are not valid in the range of $0 < K < 1$ for the proposed transformation}.
{Finally, we want to mention that  we have considered only isotropic matter distributions. The inclusion of local anisotropies into the compact configuration could reveal new insights in order to clarify at least from the theoretical point of view. So, this study will be considered elsewhere.}\\

\section*{Acknowledgments}
S. K. Maurya acknowledge that this work is carried out under TRC project-BFP/RGP/CBS/19/099 of the Sultanate of Oman. F. Tello-Ortiz thanks the financial support by the CONICYT PFCHA/DOCTORADO-NACIONAL/2019-21190856, grant Fondecyt No. 1161192, Chile and the projects ANT-1856 and SEM 18-02 at the Universidad de Antofagasta, Chile.

\appendix
\section{ The expressions for used coefficients in the above equations:}
$\Omega_{f_ 2} (x) = (1/2)\, b_1^2\, n\, \textrm{sec}^2 x\,\sin(2\,n\,x) + \cos^2 (n\,x)\, [-a_1\,b_1\, n + (a_1^2 -b_1^2)\, n^2\,\tan x - b_1^2\, \tan^3 x]$,\\
$\Omega_{f_3} (x) = 2\,\sin^2 (n\,x) [-b_1\,n + a_1\,\tan^3 x + b_1\,n^2\, \tan x\, \tan (n \,x) - n\, \tan^2 x(2 b_1 + a_1\,\tan (n\, x))] + 2\,n \,\tan x\, [\tan x\,\big(b_1 + a_1 \tan (n\,x)\big) + n\,\big(a_1 - b_1\,\tan (n\,x)\big)]$,\\
$\Omega_ {f_ 4} (x) = n\, [2\,a_1\, b_1\,\cos (2\,n\, x) + (a_1^2 + b_1^2)\,n \,\sin (2\,x) + (a_1^2 - b_1^2)\,\sin (2\, n \,x)],$\\
$\Omega_{f_ 5} (x) = \sin x\, [b_1\, \cos (n\, x) + a_1 \,\sin (n\, x)] + n\, \cos x\, [a_1\, \cos (n\,x) - b_1\, \sin (n\,x)]$,~~\\
$\Omega_ {f_ 6} (x) = 2 \cos (n\, x) \sin x [b_1 + a_1 \tan (n\, x)],$\\
$\Omega_{f_ {11}} (x) = \frac{\Omega_ {f_ 4} (x)}{[\Omega_ {f_ 5} (x)]^2}$,~~~
$\Omega_ {f_ {12}} (x) = \frac{[-5 + K + (K-1)\,\cos (2\, x)]\,\tan (x)}{\cos^2 x\, (K-1)}$ ,\\
$\Omega_ {f_ {13}} (x) =\frac{\Omega_ {f_ 6} (x)}{\Omega_ {f_ 5} (x)},~~~\Omega_ {f_ {14}} (x) =\frac{(3 - K + (K-1) \sin^2 x)}{(K - 1)\,\cos^2 x)},$ \\
$\Omega_{E_{21}}(y)= [\big(1 - m^2 + (1 + m^2) \cosh (2\,x)\big) \coth(x) \big(a_2 \cos (m\,x) + b2 \sin (m\, x)],$\\
$\Omega_ 1 (y) = [a_ 2 \cos (m\, y) + b_ 2 \sin (m\, y)]$,~~\\
$\Omega_ 2 (y)= m\, [a_ 2\, \sin (m y) - b_ 2 \, \cos (m y)],$\\ 
$\Omega_ 4 (y) =\cosh y \, \Omega_ 1 (y) + \Omega_ 2 (y)\, \sinh y,$~~\\
$\Omega_ 3 (y) = [b_ 2 m \cos (m\, y) + a_ 2 \cos (m\, y) \coth y -a_ 2 m \sin (m\, y) + b_ 2 \coth y\, \sin (m \, y)] \sinh y,$\\
$\Omega_ {f_ {21}} (y) = \frac{(3 - K + (K - 1) \cosh^2 y)\,}{(K - 1)\, \sinh^3 y\,\tanh y},~\Omega_ {f_ {22}} (y) = \frac{(1 + m^2)\, \Omega_ 1 (y)\, \Omega_ 3 (y)}{[\Omega_ 4 (y)]^2}$,~\\
$\Omega_ {f_ {23}} (y) =\frac{(m^2 +  \textrm{csch}^2 y) \Omega_1 (y) + \coth y\, \Omega_ 2 (y)}{\Omega_ 4 (y)},~~~ \Omega_ {f_ {24}} (y) =\frac{\Omega_ 3 (y)}{\Omega_ 4 (y)},$ \\
$\Omega_ {E_ {22}} (y) = \cosh^2 (m y)\,[-c_2\, d_2\,m\,\textrm{csch}^2 y + \coth y\, \big(2\, d_2^2\, m^2 + c_2^2\, (-1 + m^2) - c_2^2\,\textrm{csch}^2 y\big)],$\\
$\Omega_ {E_ {23}} (y)= (1/2)\,\textrm{csch}^2 y\,\coth y\, \sinh (m \,y)\,[1 - m^2 + (1 + m^2) \\\cosh (2\,y)]\, [2\, c_2 \cosh (m\,y) + d_2 \sinh (m\, y)]+ m [-2 c_2 \sinh^2 (m\,y) + d_2 \sinh (2\, m \,y)],$~~~
$\Omega_ 5 (y)  =  m\,\sinh y\,[d_2 \cosh (m\, y) +  c_2 \sinh (m\,y)] \\- \cosh (y) [c_2 \cosh (m\,y) + d_2 \sinh (m\, y)],$\\
$\Omega_ 6 (y)  =  \cosh (m\,y) [d_2\,m\,\coth y + c_2\,(m^2 - \textrm{csch}^2 y)]\\ + (c_ 2 \, m \, \coth y + d_ 2\,[m^2 - \textrm{csch}^2y)]\,\sinh (m\,y),$\\
$\Omega_ 7 (y) = (m^2-1)\,\sinh y\, [\cosh (m \,y) (d_ 2\, m + c_ 2 \coth (y)) + (c_ 2\, m + d_ 2 \coth y) \sinh (m\,y)]\,[c_2 \cosh(m\,y) + d_2\,\sinh(m\,y)],$~~\\
$\Omega_ 8 (y) =\cosh (m\,y) (d_ 2\,m + c_ 2 \coth y) + (c_2\,m + d_2\,\coth y) \sinh (m\,y),$\\
$\Omega_ {f_ {31}} (y) = {\Omega_6 (y)}/ {\Omega_ 5 (y)}, ~~~\Omega_ {f_ {32}} (y) = {\Omega_ 7 (y)}/{(\Omega_ 5 (y))^2}$,~~\\
$\Omega_ {f_ {33}}(y) = {\Omega_ 8 (y)}/{\Omega_ 5 (y)}$, \\
$\Omega_ {9} (y) = f_2\,(e_2 + f_2 \,y) \textrm{csch}^2y - \coth y\, [e_2^2 + 2 \,e_2 \,f_2 \,y + f_2^2\,(2 - K + y^2) + (e_2 + f_2\,y)^2\,\textrm{csch}^2 y]+(K-2) \,(e_2 + f_2\,y)^2 \,\coth^3 y,$\\
$\Omega_ {10} (y) = [3 - K + (-1 + a) \cosh (y)^2]\, \coth (y) \,\textrm{csch}^2 y,$\\
$\Omega_ {11} (y)= [(K-3) (e_2 + f_2\, y)\, \cosh y - (K-1) \,f_2\, \sinh y],$\\
$\Omega_ {12} (y) = (e_2 + f_2 \,y)\cosh y - f_2 \sinh y,$\\~
$\Omega_ {13} (y) = [2 f_2 \cosh y - (K-3) (e_2 + f_2 y) \sinh y],~\\ \Omega_{f_{41}} (y) = (e_2 + f_2\,y)\,\sinh y\,\Omega_ {11} (y)/[\Omega_ {12} (y)]^2,$\\ 
$\Omega_ {f_ {42}} (y) = \Omega_ {13} (y)/\Omega_ {12} (y),$\\
$\Omega_ {f_ {43}} (y) = \Omega_ {11} (y)/\Omega_ {12} (y).$\\
$v_{11}(x)=(a_1\, n + 3\,b_1\, n^2\, \tan x + 2\,b_1 \tan^3 x)$, \\
$v_{12}(x)=n\,\cos x\, [a_1 \cos (n\,x) - b_1 \sin (n\,x)]$,\\
$v_{13}(x)= 2 \tan x  \big(1 - \Omega_{f_{13}} (x)\big)$,\\
$v_{21}(y)=\, b_2\, m\, \cos (m\,y) + 2\,a_2\, m\, \sin (m\,y)$,\\
$v_{22}(y)=[\chi\, \Omega_ {f_{21}}(y)\,\tanh y  + (8 \pi+3 \chi )\,\Omega_ {f_ {24}} (y)]$,\\
$v_{23}(y)=[(8 \pi+3 \chi )\, \Omega_ {f_ {21}} (y)\,\tanh y  + \chi\, \Omega_ {f_ {24}} \
(y)]$,~\\
$v_{31}(y)=[2\,\coth y\, \textrm{csch} y - 3\, \Omega_ {f_ {21}} (y)] $,\\ 
$v_{32}(y)=[\chi\,\Omega_{f_{21}}(y) + (8\pi+3\chi)\,\, \Omega_{f_{33}}(y)]$,\\
~~$v_{33}(y)=[(8\pi+3\chi) \, \Omega_{f_{21}}(y) + \chi\, \Omega_{f_{33}}(y)]$,\\
$v_{25}(y)=\big(\Omega_ {f_ {22}} (y) + \Omega_ {f_ {23}} (y)\big)$,\\
$v_{41}(y)=[(e_2 +f_2 y)\, \cosh y - f_2 \sinh y]^2,\\
v_{42}(y)=[\Omega_ {f_ {41}} (y) + \Omega_ {f_ {42}} (y)].$
\\\\

\end{multicols}

\clearpage

\end{document}